\documentclass[]{aa}
\bibpunct{(}{)}{;}{a}{}{,}

\newcommand{\te}{\ensuremath{T_{\rm eff}}}

\newcommand{\bz}{\ensuremath{\langle B_z \rangle}}

\newcommand{\pv}{\ensuremath{P_V}}

\newcommand{\bs}{\ensuremath{\langle \vert B \vert \rangle}}

\newcommand{\kms}{km\,s$^{-1}$}
\newcommand{\ha}{H$\alpha$}
\newcommand{\hb}{H$\beta$}
\newcommand{\grw}{Grw\,+70$^\circ$\,8247}
\newcommand{\tpv}{20\,pc volume}

\usepackage{graphicx}
\usepackage{txfonts}

\begin{document}

   \title{Discovery of six new strongly magnetic white dwarfs\\
          in the 20 pc local population}

   \author{Stefano Bagnulo
          \inst{1}
          \and
          John D. Landstreet
          \inst{1,2}
          }

   \institute{Armagh Observatory and Planetarium, College Hill,
              Armagh BT61 9DG, Northern Ireland, UK \\
              \email{stefano.bagnulo@armagh.ac.uk,
                     john.landstreet@armagh.ac.uk}
         \and
             University of Western Ontario, Department of Physics \& Astronomy,             London, Ontario, Canada N6A 3K7\\
             \email{jlandstr@uwo.ca}
             }

   \date{Received June 2, 2020; accepted October 5, 2020}
   \titlerunning{Six new strongly magnetic white dwarfs in the 20 pc local population}
   \authorrunning{S. Bagnulo \& J. D. Landstreet}

  \abstract{
The sample of white dwarfs included in the local 20\,pc volume documents, fairly accurately, the total production of white dwarfs over roughly 10\,Gyr of stellar evolution in this part of the Milky Way Galaxy. In this sample, we have been systematically searching for magnetic white dwarfs. Here we report the discovery of six new magnetic white dwarfs, with a field strength from a few MG to about 200\,MG. Two of these stars show H lines that are split and polarised by the magnetic field. One star shows extremely weak spectral lines in intensity, to which highly polarised narrow features correspond. The three other stars have featureless flux spectra, but show continuum polarisation. These new discoveries support the view that at least 20\,\% of all white dwarfs in the local 20\,pc volume have magnetic fields, and they fully confirm the suspicion that magnetism is a common rather than a rare characteristic of white dwarfs. We discuss the level and the handedness of the continuum polarisation in the presence of a magnetic field in cool white dwarfs. We suggest that a magnetic field with a 15\,MG longitudinal component produces 1\,\% of continuum circular polarisation. We have also shown that the problem of cross-talk from linear to circular polarisation of the FORS2 instrument, used in our survey, represents an obstacle to accurate measurements of the circular polarisation of faint white dwarfs when the background is illuminated, and polarised, by the moon.
}
   \keywords{polarisation -- stars: white dwarfs -- stars: magnetic fields
                }

   \maketitle
%
%________________________________________________________________

\section{Introduction}\label{Sect_Introduction}
White dwarfs (WDs) are the end product of at least 90\,\% of stellar evolution in the part of the Milky Way Galaxy that is inhabited by the Sun. All single stars with initial main sequence masses $\la 8\,M_\odot$ end their evolutionary lives by collapsing as WDs. The same fate befalls stars in wide binary systems. Close binary systems will interact, exchange mass, and end in a wider variety of possible end states, but many of these systems produce single or double WDs. A volume-limited sample of the current population of WDs near the Sun is reasonably representative of the products of the most completed stellar evolution at this distance from the Galactic Centre. 

Because the WD cooling rate is governed by rather simple physics \citep{Mest52}, we can determine approximately when each WD formed; and from the mass determination, assuming that the WD has evolved as a single star, we can determine the approximate mass of the progenitor, using an empirically established initial-final mass relationship \citep[see, for instance,][]{Cumetal18}. Thus, a detailed study of the nearby sample of WDs provides valuable information about the local Galactic history of star formation and evolution. In addition, particular observed features of individual WDs, such as surface chemistry, the presence of one or more companions, rotation, the internal structure as deduced from pulsations, and magnetic fields, furnish important constraints on the details of stellar evolution of the specific progenitor of each WD. 

The basic sample of WDs that has been most intensely studied to obtain the kinds of information that a volume-limited sample contains is the sample of all WDs known to be within 20\,pc or less from the Sun. This sample, as refined using Data Release 2 of the {\it Gaia} astrometric survey (Gaia Collaboration \citeyear{Gaia18}), contains about 150\,WDs \citep{Holbetal16,Holletal18} and is thought to be nearly complete (possibly as many as three or four WDs have yet to be discovered). 

One of the characteristics that distinguishes some WDs from otherwise similar stars, is the presence, in a fraction of them, of a detectable magnetic field organised at a large scale over the stellar surface. These observed fields do not exhibit detectable secular changes on a readily observable time-scale, and they are usually thought to be 'fossil' fields that are present, or generated, in the precursor of the WD during an earlier stage of evolution, and they are retained during the final WD stage of evolution because of the extraordinarily slow pace of Ohmic decay, which has a time-scale of the order of 10\,Gyr \citep{Fontetal73}. Observed fields of WDs might be due to a magnetic flux retained from the magnetic fields observed in the magnetic Ap and Bp stars of the main sequence, which is amplified by the contraction of the stellar remnant from its main sequence size \citep{Angeetal81}. Alternatively, observed WD fields might have been produced by a dynamo action in the core of the star during the preceding red giant or AGB evolution stages, and they were retained by a flux conservation into the WD stage \citep{Cantetal16}. A third possibility is that the magnetic WDs (MWDs) are the result of a binary evolution of two stars which evolve to a common envelope stage during which a dynamo acts to create a magnetic field in the resulting WD \citep{Toutetal08}, or in which a close binary of two white dwarfs merges \citep{Garcetal12}. Finally, observed fields may be generated by a dynamo acting in the degenerate CO core of an already formed and cooling WD, which is driven by the crystallisation and settling of O to the centre of the core \citep{Iseretal17}. More than one of these mechanisms may act to produce the observed MWDs. Detecting and studying the magnetic field of individual WDs may provide insight into the actual actions of these, or other, processes. Surveying the WDs of a local volume-limited sample is the best way to acquire an unbiased statistical characterisation of the magnetism in the final stages of stellar evolution.

Searches for and study of magnetism among the WDs of the \tpv\ has been going on since the discovery of the first MWD. This work has been carried out by a large number of participants, using a variety of methods, mostly involving either observations of Zeeman splitting of spectral lines in optical spectra, or by employing polarimetry or spectropolarimetry to detect extremely large or extremely small magnetic fields \citep{Ferretal15}. These studies have finally established that at least about 20\,\% of WDs host magnetic fields \citep{LandBagn19b}. The global strengths $B$ of these fields range from a few kGauss (1\,kG = 0.1 Tesla) up to about 1000\,MG.
The observed fields seem to be organised on a large scale over the stellar surface.

The discovery of fields in the range of 1--2\,MG up to 100\,MG among the \tpv\ WDs, which often happens as a byproduct of low resolution classification spectroscopy, has been fairly complete for some years. In contrast, the detection of weaker or much stronger fields, which requires dedicated observations by high-resolution spectroscopy or by spectropolarimetry, has been done successfully but rather sporadically \citep[e.g.][]{SchmSmit95,Aznaetal04,Joretal07,Kawketal07,Koesetal09,KawkVenn14}. During the last five years we have been carrying out a systematic spectropolarimetric survey of the nearest WDs, and have discovered several MWDs with comparatively weak fields \citep{Landetal12,Landetal15,Landetal16,BagnLand18,BagnLand19b,LandBagn19a,LandBagn19b}.
We are now are fairly close to having complete data on the \tpv.  In this paper we report the discovery of magnetic fields in six WDs of the \tpv.

This paper is organised as follows. In Sect.~\ref{Sect_Minus} we comment on the sign of the observed polarisation. In Sect.~\ref{Sect_Estimates} we discuss how to estimate the mean longitudinal magnetic field from the measurements of circular polarisation in the continuum. In Sect.~\ref{Sect_Observations} we present the instrument and instrument settings, and summarise our observing strategy and data reduction; in Sect.~\ref{Sect_Results} we present our results on individual stars, and in Sect.~\ref{Sect_Policont_After} we critically review the reliability of our detection of polarisation in the continuum. In the final two sections we discuss the significance of the newly discovered fields, and summarise the results of this paper.

\section{On the handedness of circular polarisation}\label{Sect_Minus}
In one of the early works about polarimetry of WDs, \citet{LandAnge71} explicitly defined that circular polarisation is positive when the electric field vector, in a fixed plane perpendicular to the direction of propagation of the light, is seen to rotate counterclockwise to an observer facing the source. We call this Left-handed Circular polarisation (LCP), while Right-handed Circular polarisation (RCP) describes the situation in which the electric field vector is seen rotating clockwise. The definition $V$=LCP$-$RCP was followed until the end of the last century in most of the publications reporting polarimetric observations of MWDs \citep[e.g.][]{Angeetal81,SchmSmit95,Putn97}, of magnetic Ap stars \citep[e.g.][]{BorLan80} and of other stars in the HR diagram \citep[e.g.][]{Donaetal97}.  During the past two decades, using the ISIS, MuSiCoS, FORS, ESPaDOnS and Narval instruments, either deliberately or by chance, several groups have adopted the opposite definition, that is, $V$=RCP$-$LCP (a long reference list starts with the observations of magnetic Ap stars by \citeauthor{Wadetal00a} \citeyear{Wadetal00a}). This definition conforms, for instance, to \citet{LanLan04}, who justified their choice by linking the handedness of the polarisation to the helicity (or spin) of photons (see their Sect. 4.4 and App. 3); the same definition was adopted in various numerical simulations of polarised spectra from magnetic stars \citep[see, for instance., the comparison of three radiative transfer codes in magnetic atmospheres by][]{Wadetal01}.

It is not up to us to decide which definition is best to adopt for $V$, but we remind readers that it will be necessary to keep the sign ambiguity in mind and possibly to change the sign of some observations for comparison purposes, as was done for \grw\ by \citet{BagnLand19a}. In the work reported here, and all our recent publications on MWDs, we have followed the convention that $V$=RCP--LCP.

\section{Estimating the mean longitudinal magnetic field from circular polarisation measurements of the continuum}\label{Sect_Estimates}

The initial discovery of white dwarf magnetism succeeded as a result of James Kemp's idea that radiation from electrons embedded in a strong magnetic field should produce circularly polarised continuum light \citep{Kemp70, Kempetal70a}, and his subsequent follow-up of this idea, with key suggestions from George Preston and one of the authors \citep[see][]{Land20}, by an observational search for polarisation of DC white dwarfs. This survey revealed strong circular polarisation in the white dwarf \grw\ = WD\,1900+705 \citep{Kempetal70}, and thus this star became the first MWD discovered. 

This discovery stimulated extensive searches for further MWDs using broad-band circular polarimetry. During the next five years, six further MWDs were discovered using this technique \citep{AngeLand71,LandAnge71,AngeLand72,Swedetal74,Angeetal74, Angeetal75}. It was only in 1974 and 1975 that the first examples of clear Zeeman splitting of spectral lines observable in WD flux spectra due to fields of order 10\,MG were identified \citep{Angeetal74b,Liebetal75,WickBess76,Liebetal77}. Because of the relative prominence of broad-band circular polarimetry, and later spectropolarimetry, in the early discovery work, significant effort was expended on developing methods of interpreting the broad-band polarisation, and using the observations to estimate line-of-sight field strengths \bz\ in the newly discovered MWDs.

The interpretation of observed continuum polarisation involves two main aspects. First, the differences between emissivity (and absorptivity) of atmospheric constituents (free electrons, H, He, H$^-$, He$^-$) of right and left circularly polarised radiation produced in a magnetic field needs to be determined or estimated. Secondly, the different polarised absorption processes need to be inserted into a radiative transfer code to compute how absorption difference at the atomic level result in various amounts of observed stellar polarisation as a function of \bz, \te\ and atmospheric composition. 

Kemp's result \citep[as corrected by him, see][]{Kemp70,Kemp77} for the polarisation of a heated, radiating medium due to brehmsstrahlung by free electrons is that
\begin{equation}
    \vert V/I \vert= \left| \frac{j_{\rm RH}- j_{\rm LH}}{j_{\rm RH}+ j_{\rm LH}} \right|
         = \left| \alpha_1 \frac{\nu_{\rm L}}{\nu} \right|
         = \left| \alpha_1 \frac{e B \lambda \cos(\theta)}{4 \pi m c^2} \right|
\label{Eq_Kemp}
\end{equation}
where $V$ and $I$ are Stokes $V$ and $I$, $j_{\rm RH}$ ($j_{\rm   LH}$) is the emissivity in right-hand (left-hand) circularly polarised light; $\alpha_1$ is a constant of order 1 that depends  on the emission  mechanism;  $\nu$ is  the  frequency of  the radiation of wavelength $\lambda$;  $B \cos(\theta)$ is the component of the  magnetic field along  the line of sight; and  $\nu_{\rm L} = eB/4 \pi m c$ is the Larmor  frequency (in cgs Gaussian units). In Eq.~(\ref{Eq_Kemp}) we have used the absolute value ($\vert \ldots \vert$) to avoid the problem of sign ambiguity that was discussed in Sect.~\ref{Sect_Minus}.

\citet{Ship71} noted that Kemp’s prediction, that the circular polarisation should increase linearly with wavelength (see Eq.\,\ref{Eq_Kemp}), is inconsistent with observations, and suggested that combining Kemp’s polarisation theory with radiative transfer would lead to a better agreement with observations. The effect of radiative transfer on the polarised emissions is to make the physical depth to the optical depth from which radiation emerges in right and left polarised light slightly different for the two polarisations, leading to a difference between the emergent fluxes in the two polarisations. \citet{Ship71} estimated the difference between the opacities in right and left polarisation as 
\begin{equation}
\left| \frac{\kappa_{\rm RH}-\kappa_{\rm LH}}{\kappa_{\rm RH}+\kappa_{\rm LH}} \right| =  \left| 4 \frac{e B \lambda \cos(\theta)}{4 \pi m c^2} \right| \;.
\label{Eq_Shipman}
\end{equation}
Shipman computed emergent flux from a model atmosphere with normal opacities, and fluxes from a second atmosphere with all opacities increased according to Eq.\,(\ref{Eq_Shipman}), and took the flux difference as the continuum polarisation of \grw. He found a reasonable fit to the observed circular polarisation spectrum, and deduced a 
longitudinal field strength between  6 and 10\,MG, depending on the atmospheric composition. This led to the conclusion that a 2-3\,MG magnetic field parallel to the line of sight would lead to 1\,\% of circular polarisation.
In parallel, Shipman developed a simple analytical model that predicted the level of continuum polarisation and its wavelength dependence. This analytical model was improved and applied repeatedly during the 1970s, as discussed below. 

\citet{AngeLand74} extended and developed Shipman's analytical model, and used it to interpret the observed  smooth, roughly constant observed continuum polarisation (about 0.8\,\% around 5000\,\AA)  of the carbon-rich star G99-37 = WD\,0548$-$001 ($\te = 6080$\,K), assuming  
$\bz = 3.6$\,MG as determined using the molecular Zeeman effect in the CH G-band. 

A quantitative test of field strength deduced using the extended Shipman method was reported by \citet{Liebetal75}. For the star G99-47 = WD\,0553+053 ($\te = 5790$\,K), a well-determined,  almost constant continuum polarisation of about 0.35\,\% yielded an estimate of $\bz = 1.2$\,MG (assuming an H-rich atmosphere). However, using spectropolarimetry of the H$\alpha$ line, \citet{Liebetal75} made a direct measurement of $\bz = 5.6$\,MG. This suggested that (at least at this \te) the best available method of estimating the longitudinal field using continuum polarisation substantially underestimated the true field value. 

The calibration of the continuum polarisation -- \bz\ relationship was reviewed by \citet{Angeetal81}, who discussed the usefulness of a simple relationship of the form
%--------------------------------------
\begin{equation}
   B \cos \theta = \bz  \approx  \gamma\ (V/I). 
\label{Eq_Land}
\end{equation}
%--------------------------------------
They concluded that $\gamma \approx 5$\,MG per \% polarisation (within a factor of 2) over the range between $\te = 5000$ and 25000\,K (see their Fig.~1). It is interesting to note that this estimate is qualitatively consistent with that of \citet{ChiSil76} who, from the analysis of the continuum polarisation based on a proper (but still simplified) treatment of the radiative transfer in a magnetic atmosphere, derived a 5.3\,MG field from the observed 0.8\,\% polarisation of G\,99-37, and 13\,MG from the 2\,\% polarisation of G\,227-35.

Attention during the 1980s focussed on exploiting the major efforts to compute the atomic spectra of neutral H and He in really large magnetic fields ($B > 100$\,MG), far beyond the linear and quadratic Zeeman effect regimes. Computations of atomic energy levels, wavelengths of spectral lines,  and transition probabilities as a function of field strength were provided particularly by groups at Louisiana State University \citep{HenrOcon85} and the Eberhard Karl University of T\"{u}bingen
\citep{Forsetal84,Roesetal84}. It was found that these data finally made it possible to obtain computed intensity spectra of \grw\ that showed structure similar to that observed in the flux spectrum of this MWD \citep{Angeetal85,Wunnetal85}. In parallel, much important work was done by the group of Dayal Wickramasinghe to develop real radiative transfer codes that could be used to model the intensity and polarisation spectra of MWDs \citep{MartWick79} without relying on the very simple approximations used during the 1970s.  A decade later Stefan Jordan developed powerful new programmes to compute detailed atomic line flux and polarisation spectra \citep{Jord92}. All these efforts led to detailed and reasonably realistic models of the flux (Stokes $I$) from several individual MWDs, and eventually grids of spectral flux models were produced to estimate the field strengths \bs\ in the many strong-field WDs discovered in the spectra obtained by the SDSS project \citep[e.g.][]{Kepletal13}. 

In principle, there is now a large body of knowledge and  powerful computational tools that we could exploit to model the flux and polarisation signatures in the WDs in which we have discovered evidence of magnetic fields. However, a major computational effort (usually guided by spectral line features in Stokes $I$) is required to establish even a plausible model for any single WD, and such an effort is beyond the scope of this work. Therefore, we try to extract from the literature some guidance about how to estimate usefully the likely magnetic fields underlying the polarised spectra that we observe. Unfortunately, the recent important advances in computational capability have been applied primarily to WDs that are hotter than the sample that will be discussed in this paper, that is, to objects in which the flux spectrum contains signatures of spectral lines, providing powerful constraints on composition and approximate field strength. Recent work has not specifically addressed the need for a simple method with which to estimate the order of magnitude of the field strength on the WD surface from the magnitude of the observed continuum circular polarisation in cool magnetic stars. 

\citet{Angeetal81} provide an approximate calibration of $\gamma$ in Eq.\,(\ref{Eq_Land}) as a function of \te\ and composition. However, this calibration is based on a severe simplification of the radiative transfer problem, and is probably of quite limited accuracy. In this paper we revisit this problem.  What we ideally need for calibration is a selection of cool MWDs for which reasonably accurate continuum polarisation measurements (preferably as a function of wavelength) exist, and for which measurements and modelling have provided values of the mean surface field \bs\ and particularly of the mean longitudinal field \bz, because we suppose that the level of continuum circular polarisation is determined essentially by \bz\ rather than \bs. The sample of possible calibrating stars is actually still, after 40 years, quite limited, mainly because only a very small fraction of the known MWDs have been observed with filter polarimetry or spectropolarimetry.

\subsection{The sign of the magnetic field from the continuum polarisation}\label{Sect_Sign}
Preliminary to our calibration exercise, we wish to scrutinise the relationship between the sign of the continuum polarisation and the sign of the magnetic field that is responsible for it. In the past, it has been assumed that { \bz\ has the same sign as the continuum polarisation defined as LCP$-$RCP \citep[see, for instance Table~2 of][]{Angeetal81}. Consistently with \citet{Angeetal81},} \citet{Putn97}, in Sect.~5.2, states explicitly that the negative sign of the polarisation observed in G\,111-49 points to a negative value of the "effective field" (which we presume is the same quantity as \bz). { However, both observational evidence and theoretical calculations point to a relationship with the opposite sign when $V$ is taken to be LCP--RCP.

As observational evidence, we first discuss four cases from the literature, G\,99$-$47=WD\,0553$+$053, G\,227--35=WD\,1829+547. G\,111$-$49=WD\,0756$+$437, and Grw+70\degr\,8287 = WD\,1900$+$705.

 The data presented by \citet{Liebetal75} and \citet{PutnJord95} for G\,99$-$47 show a polarised continuum with  LCP>RCP (hence positive in the 'old' convention), consistent with several broadband values ($\sim 0.4$\,\%) reported in the discovery paper by \citet{AngeLand72}. Since the weak continuum polarisation of G\,99--47 was independently measured by three groups, we consider this detection to be very reliable. Both spectropolarimetric datasets also agree that the blue $\sigma$ component of \ha\ has LCP>RCP. In the Zeeman regime, the blue $\sigma$ components of an absorption line have LCP>RCP if the longitudinal field is negative, therefore our conclusion is that if one uses the convention adopted by both \citet{Liebetal75} and by \citet{PutnJord95}, that is, $V$=LCP$-$RCP, a positive continuum polarisation corresponds to a negative value of \bz. In the current convention of $V$=RCP$-$LCP, positive continuum polarisation would correspond to positive field.
 
 In addition to G\,99--47, in which the sign of the field \bz\ is clear from the easily visible full Zeeman signature, we know of three stars with much stronger fields, in which the polarisation in the continuum is firmly detected, and in which the sign of the field can be deduced from the signs of single magnetic subcomponents, particularly that due to the strong 2p-1--3d-2 transition. Spectra of G\,227--35, G\,111--49, and \grw\ are all displayed in Fig.~3 of \citet{Putn95}. In each of these MWDs, the 2p-1--3d-2 transition extends towards positive polarisation values relative to the local (negative) continuum, and so, if any blue $\sigma$ subcomponents were visible in the spectra, they would extend towards values more negative than the local continuum. This behaviour of the $\sigma$ components is interpreted by positive \bz\ values for all three stars, while the sign of the continuum polarisation measured with the old convention $V$=LCP-RCP is negative. In the current polarisation convention $V$=RCP-LCP, again we would find that the sign of \bz\ is the same as the continuum polarisation. 
 
 Finally, in Sect.~\ref{Sect_WDSeventeen} we show that also our new observations of WD\,1703$-$267 suggest that a positive \bz\ is responsible for RCP > LCP, consistently with the four cases from the previous literature. Our conclusion is that observational evidence suggests that if a magnetic field is oriented towards the observer, the electric field vector associated to the continuum of the incoming radiation is seen rotating clockwise in a fixed plane looking at the source.  

Regarding theoretical results, we note that the modelling of G\,99$-$47 by \citet{PutnJord95} predicts that a magnetic field with a negative longitudinal component (as deduced from the Stokes $V$ profiles of spectral lines), produces a continuum polarisation with RCP < LCP (see bottom panel of their Fig.~2). Numerical simulations by \citet{AchiWick89} predict that a field with positive longitudinal component has RCP > LCP  (see their Fig.~2).  

In conclusion, if one defines Stokes $V$=LCP$-$RCP, a positive \bz\ produces a negative polarisation in the continuum, and the $\gamma$ value in Eq.~(\ref{Eq_Land}) should be negative. { Conversely}, if one defines $V$=RCP$-$LCP (our current convention), a positive \bz\ produces a positive polarisation in the continuum, and Eq.~(\ref{Eq_Land}) should have a positive value for $\gamma$. To minimise ambiguity, in the following we refer to the absolute value $|\gamma|$.
Regardless of the definition of $V$, we observe in these examples that if a magnetic field is strong enough to produce continuum polarisation and the Stokes $V$ profile of spectral lines can be observed, then the continuum polarisation has the same sign as the difference between the $V$ value of the blue $\sigma$ components of spectral lines and the underlying continuum, and opposite to the sign of the difference between the $V$ value of the red $\sigma$ components and the continuum.}

\subsection{Calibrations from individual MWDs}
In the following we discuss individual calibrators that will help to estimate the mean longitudinal magnetic field from the circular polarisation in the continuum. Suitable MWDs for this test include WD\,0553+053, WD\,0548--001, WD\,1015+014, WD\,1829+547, and WD\,0756+427, which are discussed below. For a number of other cool MWDs such as WD\,0011--134, WD\,0503--174, WD\,1026+117, and WD\,1309+853, continuum polarisation measurements exist and the field strengths \bs\ have been estimated \citep{Putn97}, but for these MWDs $V(\lambda)/I$ has not been measured with high enough precision to be useful for the calibration we need. Furthermore, frequently the published data are available only in graphical form, so we cannot obtain higher precision by binning in wavelength.

\subsubsection{WD\,0553+053 = G\,99$-$47}
This star, with  $\te = 5790$\,K \citep{Limoetal15}, may be the most useful calibrator as is the closest in \te\ to the stars analysed in the next section. It was already used as a calibrator by \citet{Liebetal75}, yelding $|\gamma| \sim 16$\,MG per \%\ of polarisation. It is useful to revisit this calibration using the  $I$ and $V/I$ spectra obtained by \citet{PutnJord95}, because these data have considerably higher {\it S/N} than those used by \citet{Liebetal75}. We have digitised these more recent data, and measured them. Using the tables of \citet{SchiWunn14} with the $I$ spectrum, we estimate $\bs = 14.0 \pm 0.5$\,MG. Computing the mean separation of \ha\ as seen in right and left polarised spectra, we find $\bz = -6.2 \pm 1$\,MG. These values (as well as the actual spectral data) are very similar to those reported by \citet{Liebetal75}. Our negative value of \bz\ agrees with the statement by \citet{Liebetal75} that the field of G\,99-47 points away from the observer.  Outside of the Balmer line regions, the data can plausibly be fitted with a constant or nearly constant value $\sim 0.35$\,\% around 5000\,\AA. Combining this value of $V/I$ with the measured value of \bz, we find $|\gamma| \approx 18 \pm 2$\,MG per \% circular polarisation. In absolute value, this is considerably larger value than that deduced from model computations by \citet{Angeetal81}, but consistent with the earlier empirical result of \citet{Liebetal75}. 

\subsubsection{WD\,0548--001 = G\,99--37} This WD is a cool star \citep[with $\te = 6200$\,K and a He-rich atmosphere according to][]{Vornetal10}. It shows fairly strong Swan bands of C$_2$ and a strong CH G-band. It is within the \tpv. The G-band, produced by an asymmetric molecule, has reasonably strong Zeeman splitting and polarisation, studied by \citet{AngeLand74} and more recently by \citet{Berdetal07} and \citet{Vornetal10}. The star also displays significant continuum circular polarisation of about 0.8\,\% that varies weakly with wavelength outside the strong Zeeman signature observed in the G-band.  Both recent studies agree that the value of \bz\ is about 7.3\,MG, based on detailed modelling of the polarisation of the G-band, together with simplified radiative transfer. The value of \bs\ is unknown.  Taking these data at face value, this MWD suggests a value of $| \gamma |$ of about $9 \pm 3$\,MG/\% polarisation. 

\subsubsection{WD\,1015+014 = PG\,1015+014} This is an MWD with $\te \approx 10000$\,K and an apparently H-rich atmosphere. Its field has been observed and modelled by \citet{WickCrop88} and by \citet{Euchetal06}, who fit both $I$ and $V/I$ spectra, and  concluded that much of the visible magnetic flux on the stellar surface has strength of about $\bs \approx 75$\,MG. Broadband $V/I$ for this star is time--variable and ranges from an average of about +1.1\,\% to $-1.35$\,\% \citep{SchmNors91}. No direct measurement of \bz\ is available, although the models of \citet{Euchetal06} suggest large regions of a single polarity predominate at some phases. If we guess that rotation provides a view of much of the magnetic field, and that the strongest projected field \bz\ is about 1/3 of \bs\ (typical for the dipole component of the field), or about $20-30$\,MG, corresponding to $\vert V/I \vert \approx 1.35$\,\%, we find $| \gamma | = 19 \pm 4$\,MG per \%\ polarisation.

\subsubsection{WD\,1829+547 = G\,227--35} Circular spectropolarimetry of this WD has been published by \citet{Angeetal75} and \citet{PutnJord95}. No significant linear polarisation is detected. The star is a cool ($\te \approx 6300$\,K) large-field MWD, with circular polarisation  that varies rather strongly with wavelength, reaching $\vert V/I \vert \approx 4$\,\% around 4500\,\AA, dropping to near $1$\,\%, and then gradually rising from 6000\,\AA\ into the infrared. 
The two sets of $I$ and $V/I$ spectra are very similar, and suggest lack of variation on a long time-scale.  \citet{PutnJord95} have modelled the available data. From the presence of a strong feature in the polarisation spectrum at about 7450\,\AA, thought to be a stationary component of \ha\ \citep{Coheetal93}, the surface composition is deduced to be H-rich, and the value of \bs\ is found to be about 120\,MG. No estimate of \bz\ is available. However, the absence of detectable linear polarisation suggests that the line of sight to the WD is roughly aligned with the magnetic axis, which could lead to a mean longitudinal field of $\sim 30 - 50$\,MG  \citep[a substantially higher value than the 13\,MG predicted by][]{ChiSil76}. If this is assumption is correct, the observed polarisation of roughly $3$\,\% (in absolute value) is consistent with $\vert \gamma \vert = 13 \pm 4$\,MG per \%\ circular polarisation.

\subsubsection{WD\,0756+427 = G\,111--49.} Strong continuum circular polarisation, varying with wavelength and rising to about $8$\,\%\ around 5500\,\AA, was discovered in the cool WD ($\te \approx 7000$\,K)  G\,111--49 by \citet{Putn95,Putn97}. The observed spectrum includes three prominent line- or band-like polarisation features longward of 6700\,\AA, and the red polarisation spectrum qualitatively resembles that of \grw. The spectral features are interpreted by Putney as strong, stationary \ha\ components, implying a field of roughly $\bs \approx 200 - 250$\,MG, so this MWD is a DA. The presence of both the extraordinarily high level of continuum circular polarisation and of strong circular polarisation line features suggests that the dipole component of the stellar field is nearly parallel to the line of sight \citep{Putn95}, so that we could reasonably guess that the value of \bz\ may be as large as 80\,MG. If this value is related to a wavelength-averaged circular polarisation of about 5\,\%, we find that $\vert \gamma \vert \simeq 16$\,MG per \%\ polarisation; if we guess a lower value of \bz\ and compare it to the peak continuum polarisation around $8$\,\%, $\vert \gamma \vert \simeq 10$\,MG per \%. In any case the range of plausible $| \gamma |$ values for this MWD is within the range of the values estimated above.

\subsubsection{Summary}
Overall, the examples above consistently suggest that in Eq.~(\ref{Eq_Land}) we should adopt $\vert \gamma \vert \sim 15$\,MG per 1\,\%\ continuum circular polarisation, using a negative value to interpret older polarimetric measurements as explained above, and a positive value to interpret our own spectra. It seems probable that $| \gamma |$ depends on \te\ and probably on composition, so we do not expect Eq.\,(\ref{Eq_Land}) to provide much better accuracy than perhaps a factor of 2, but the mean value of this coefficient is substantially larger than the value accepted by \citet{Angeetal81}.

We recall that, for fields with roughly dipolar topology, typically $\vert\bz\vert \leq 0.5 \bs$ \citep[e.g.][]{Hensetal77}. Thus the observed degree of circular polarisation should provide a rough lower limit to the field \bs, which we estimate as $\bs \ga 30$\,MG per 1\,\% of continuum circular polarisation. 

\section{New observations}\label{Sect_Observations}
During the past few years we have been using the Cassegrain FORS2 spectropolarimeter instrument mounted on UT1 of the ESO VLT to observe, or re-observe, all southern \tpv\ WDs not known to be magnetic stars, but for which literature data did not allow us to place a tight upper limit on the strength of a possible magnetic field. This section summarises the instrument settings, observing strategy, and data reduction procedures used during this study.

\subsection{Instrument settings}\label{Sect_Settings}
The FORS2 instrument allows one to carry out single-order spectropolarimetry with a variety of grisms, covering various wavelength windows between the UV atmospheric cutoff and the near IR. For WDs with strong atomic spectral lines (DA, DB or DZ WDs), circular spectropolarimetry of one or several strong lines allows us to detect, using the polarisation due to the Zeeman effect, a mean line-of-sight field component \bz\ as weak as a few hundred G \citep{BagnLand18}. For WDs of known spectral type and approximate temperature, we were able to choose grisms that best sample the available spectral lines, but for newly recognised WDs that did not yet have spectral classification, we mostly used grism 1200B, which provides a spectral window from 3700\,\AA\ to 5100\,\AA, allowing us to use possible spectral lines of H (H$\beta$ and higher), He, Ca, Mg, or Fe to detect magnetic fields. With a 1\arcsec\ slit width, grism 1200B provides a spectral resolution of $\sim 1400$. For one star we have also employed grism 600B, for wider spectral coverage (3600\,\AA\ to 6200\,\AA). With slit width set to 1\arcsec, spectral resolution was $\sim 780$.

For stars that were previously classified as DC, and are thus without significant spectral lines, we searched for broad-band continuum circular polarisation using the lower resolution grism 300V. With this grism we observe a spectral window between about 3500 and 9300\,\AA; setting the slit width to 1.2\arcsec, we obtain a spectral resolution of $\sim 360$. Spectra can be heavily rebinned without loss of significant information, and the detection limit of circular polarisation is determined by instrumental polarisation rather than {\it S/N} even on the faintest targets of our list. Instrumental polarisation and field detection limits will be discussed in Sect.~\ref{Sect_Policont_After}. Although it does not allow us to discover kG-level fields in DC WDs, spectropolarimetry of the continuum is the only search technique available for line-free WDs at present.  

For the three observations of two stars that were obtained between September 26 and 30, 2019, the blue-optimised E2V CCD was used; for the remaining eight observations of the other four stars, the MIT CCD, optimised for the red part of the spectrum, was employed. For the three observations obtained with the E2V CCD we used the 1200B grism; in this configuration, a blue-optimised CCD has a significant advantage compared to the MIT CCD.

\subsection{Observing strategy and data reduction}
The observing strategy (that is, the use of the beam swapping techniques to minimise instrument contributions) and data reduction are fully described in \citet{BagnLand18} and references therein. In particular, \citet{BagnLand18} have focussed on the detection of signals of circular polarisation in spectral lines, and their interpretation in terms of mean longitudinal magnetic field (for more details see also \citeauthor{Bagetal12} \citeyear{Bagetal12} and \citeauthor{Bagetal13} \citeyear{Bagetal13}). These considerations will not be repeated here. However, to interpret the observations presented in this paper we need to add a critical discussion of the polarimetric capabilities of FORS2 in the continuum, and about the interpretation of the circular polarisation in the continuum in terms of magnetic field. The discussion about the instrument polarimetric capabilities in the continuum (and in particular, instrumental polarisation) will be presented in Sect.~\ref{Sect_Policont_After}.% In Sect.~\ref{Sect_Estimates} we have discussed the interpretation of the signal of continuum circular polarisation in terms of magnetic field.

\subsection{The newly discovered magnetic stars}
%%%%%%%%%%%%%%%%%%%%%%%%
\begin{table*}
\caption{Properties of newly discovered MWDs} % title of Table
\label{Tab_new_mwds}      % is used to refer this table in the text
\centering                          % used for centering table
\begin{tabular}{l l l l l l l}        
\hline\hline                 % inserts double horizontal lines
WD Name                &WD\,0004$+$122     & WD\,0708$-$670    &WD\,0810$-$353    &WD\,1315$-$781    &WD\,1703$-$267    &WD\,2049$-$253 \\       
Alt Name               & LP 464-57         & SCR J0708-6706    & UPM J0812-3529   &  LAWD 45         &  Gaia DR2 only   & Gaia DR2 only  \\  
%%%& .                    &                   & .          &              &   \\
\hline
Spectral class            & DCH               & DCH               &  DAH    	  &  DAH             &  DAH             &  DC(H?)         \\
V                      &                   & 16.22             & 14.47	          &  16.16           &                  &              \\	     
G                      & 16.26             & 15.97             & 14.36    	  &  15.97           & 15.02            &  16.04       \\
$\alpha$ (J2000)       & 00 07 20.55       & 07 08 52.28       & 08 12 26.98      &   13 19 25.56    & 17 06 41.33      &  20 52 13.51  \\
$\delta$ (J2000)       & +12 30 21.16      &$-$67 06 31.43     &$-$35 29 43.8     & $-$78 23 28.17   & $-$26 43 36.2    &$-$25 04 17.5  \\
$\pi$ (mas)            & $57.31 \pm 0.11$  & $59.02 \pm 0.04$  & $89.52 \pm 0.02$ & $51.84 \pm 0.04$ & $76.62 \pm 0.06$ &  $55.65 \pm $ 0.06  \\
\te\,(K)               & $4885 \pm 45$     & $5020 \pm 80$     & $6093$           & $5619 \pm 193$   & $6168$           &  $4895$  \\
$\log g$ (cgs)         & $8.09 \pm 0.02$   & $7.99 \pm 0.04$   & $8.09$           & $8.17 \pm 0.02$  & $8.35$           &  7.84  \\    
Mass ($M_\odot$)    & $0.625 \pm 0.014$    & $0.56 \pm 0.03$   & $0.63$           & $0.69 \pm 0.02$  & $0.82$           &  0.48  \\
Age (Gyr)              & 6.45              & $5.47 \pm 0.34$   & 2.7              & $4.39$           & 4.2              &   4.4 \\
Reference              & 1, 2              & 2, 5              & 2, 3, 6          & 2, 4             & 2, 3, 6          &  2, 3, 6\\
\hline                        % inserts single horizontal line
\end{tabular}
\tablefoot{Key to references: (1) \citet{Blouetal19}; (2) Gaia Collaboration (\citeyear{Gaia18}); (3) \citet{Gentetal19}; (4) \citet{Giametal12}; (5) \citet{Subaetal17}; (6)  http://www.astro.umontreal.ca/\~{}bergeron/CoolingModels.  Spectral classification is based on this work.  }
\end{table*}

%%%%%%%%%%%%%%%%%%%%%%%%
%%%%%%%%%%%%%%%%%%%%%%%%
\begin{table*}
\caption{Log of spectropolarimetric observations of new MWDs with FORS2}
\label{Tab_journal}
    \centering
    \begin{tabular}{l l l l r r c c r@{\,$\pm$}l r@{\,$\pm$\,}l}
    \hline\hline
    Star      & Date       &  time &   MJD     & exp time & Grism   &{\it S/N} & $V/I$  &\multicolumn{2}{c}{\bs} &\multicolumn{2}{c}{\bz} \\
              &            &  (UT) &           & (s)      &         &(\AA$^{-1}$)& (\%) & \multicolumn{2}{c}{(MG)}&\multicolumn{2}{c}{(MG)} \\
    \hline
WD\,0004$+$122& 2019-10-07 & 01:17 & 58763.054 & 1560 &  300V  & 125 &
$-0.5$ to $+2.0$ &\multicolumn{2}{c}{60 or 200}&\multicolumn{2}{c}{$\sim 30$}\\[2mm]
WD\,0708$-$670& 2019-10-17 & 07:20 & 58773.306 & 1560 &  300V  & 170  &  $-0.5$ to $+1.0$ &\multicolumn{2}{c}{60 or 200}&\multicolumn{2}{c}{$\sim 15$}\\  
              & 2019-12-09 & 08:02 & 58826.325 & 1560 &  300V  & 185 & &\multicolumn{2}{c}{60 or 200}&\multicolumn{2}{c}{$\sim  15$}\\[2mm]
WD\,0810$-$353& 2019-03-24 & 01:20 & 58566.058 & 2400 & 1200B  & 460 & $-1$ to $+0.5$  &\multicolumn{2}{c}{30}         &\multicolumn{2}{c}{non-zero}\\
              & 2020-01-08 & 03:03 & 58856.127 & 3680 & 1200B  & 520 & &\multicolumn{2}{c}{30}         &\multicolumn{2}{c}{non-zero}\\ 
              & 2020-01-08 & 04:18 & 58856.180 & 3600 &  600B  & 440 & &\multicolumn{2}{c}{30}         &\multicolumn{2}{c}{non-zero}\\[2mm]
WD\,1315$-$781& 2020-01-08 & 07:26 & 58856.310 & 1600 &  300V  & 95&   $-$ & 5.5 &   0.2                & $-0.06$ & 0.12     \\
              & 2020-01-13 & 07:18 & 58861.304 & 1600 &  300V  & 170  &  & 5.5 &   0.2                & $0.03 $ & 0.08     \\[2mm]
WD\,1703$-$267& 2019-09-26 & 01:01 & 58752.042 & 3040 & 1200B  & 190  &  $+0.2$ &8.0 &   0.5                & $1.9  $ & 0.2      \\
              & 2019-09-30 & 01:29 & 58756.062 & 3040 & 1200B  & 380  &  & 8.5 &   0.5                & $2.75 $ & 0.07     \\[2mm]
WD\,2049$-$253& 2019-09-29 & 03:17 & 58755.137 & 4320 & 1200B  & 365 & $+0.5$  &\multicolumn{2}{c}{$\ga 20$}&\multicolumn{2}{c}{$\sim 7$}\\
         \hline
    \end{tabular}
\end{table*}

%%%%%%%%%%%%%%%%%%%%%%%%
The objects selected in our \tpv\ spectropolarimetric survey include a number of previously classified WDs, and several previously unrecognised WDs recently identified by \citet{Holletal18}. Six of these objects show strong evidence of the presence of previously unknown magnetic fields. Table\,\ref{Tab_new_mwds} lists the WDs in which we have discovered fields, and summarises their relevant astrometric and physical properties, including spectral classification as determined from this work. 
We note that the distinction between MWDs classified with the suffixes P and H simply refers to a different field detection method, namely via polarimetric (P) or spectroscopic (H) measurements. Because this classification does not point to a physical stellar feature, and because a number of MWDs exhibit both clear magnetic line splitting and continuum circular polarisation, we prefer to abandon this distinction and adopt the suffix H for all MWDs, regardless the actual observing technique that was employed for their discovery. Table\,\ref{Tab_journal} shows the log of our observations.

\section{Results}\label{Sect_Results}
In the following we discuss individually the newly discovered, nearby MWDs. 

\subsection{DA star WD\,1315$-$781 = LAWD\,45}
%%%%%%%%%%%%%%%%%%%%%%%%%%%%%%%%%%%%%%%%%%%%%%%%%%%%%%%%%%%%%%%%%%%%%%%%%%%%%%%%%%%%
   \begin{figure}[ht]
   \centering
   \includegraphics[width=9cm,trim={0.9cm 2.3cm 0.7cm 1.0cm},clip]{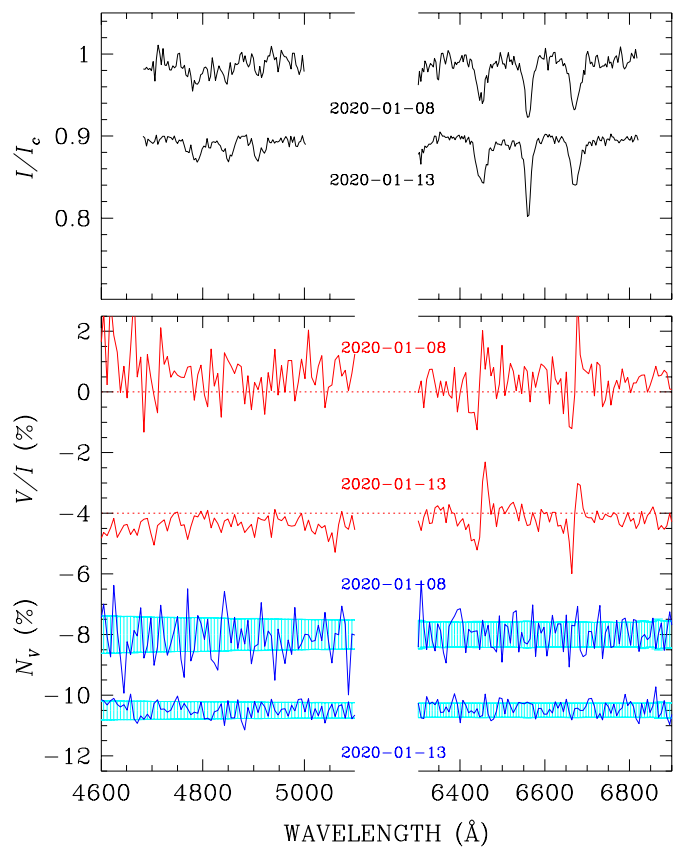}
      \caption{WD\,1315$-$781: Observed $I/I_{\rm c}$ (top panel),  $V/I$ (red curves) and $N/I$ (blue curves) spectra of WD\,1315$-$781 around H$\beta$ and H$\alpha$.
              }
         \label{Fig_wd1315}
   \end{figure}
%%%%%%%%%%%%%%%%%%%%%%%%%%%%%%%%%%%%%%%%%%%%%%%%%%%%%%%%%%%%%%%%%%%%%%%%%%%%%%%%%%%%
WD\,1315$-$781 is a well known WD, already identified from its large proper motions by \citet{Luyt49}. The basic parameters \te, $\log g$, mass, and age,  have been derived from astrometry, spectra and photometry by \citet{Subaetal07} and \citet{Giametal12}. (We note that although the parameters of this star were derived before Gaia parallaxes were available, the parallax value adopted by \citeauthor{Giametal12} \citeyear{Giametal12} is consistent with the Gaia value.) The WD is included in the \tpv\ by \citet{Holletal18} on the basis of the Gaia parallax. This star has repeatedly been classified as a DC star. However, \citet{Giametal12}, using  a low {\it S/N} spectrum, suggested very tentatively that an extremely weak H$\alpha$ might be present, and in fact that it might be split by the Zeeman effect, so that WD\,1315$-$781 could possibly be a MWD.  

Because the atmospheric composition of the WD seemed uncertain, we obtained two circularly polarised spectra using the 300V grism to cover the full visible and near IR. The star is cool enough (about 5600\,K) that even if it is H-rich, we would expect only \ha\ to be easily detectable. Indeed, this line is clearly present in both our FORS2 spectra, as shown in Fig.~\ref{Fig_wd1315}. Furthermore, it is obvious that the \ha\ line is split by the Zeeman effect, and that it shows the kind of signal in the circular polarisation $V/I$ spectrum that confirms the presence of a field. The correct classification of this star is thus DAH. 

In fact, in addition to \ha, the H$\beta$ line is also weakly present in both our spectra. In conformity to the expected Balmer decrement, the components of H$\beta$ are only about 3\,\% deep, compared to about 10\,\% for the $\pi$ component of \ha. The H$\beta$ line is split in a very similar way to \ha, and fully confirms the detection of a field. 

Consider first the more recent spectrum (shown in Fig.~\ref{Fig_wd1315}) which has significantly higher {\it S/N} per \AA\ than the earlier spectrum (170 vs. 95 per \AA). From the splitting of \ha\  \citep[see][]{SchiWunn14}, we deduce a mean field modulus \bs\ of about 5.5\,MG. Comparing the widths of the $\sigma$ components to that of the $\pi$ component, it is clear that the field on the visible hemisphere mostly varies by less than about $\pm 0.5$\,MG relative to the mean value. The $I$ profile of the weak H$\beta$ confirms these values at a considerably lower S/N level. 

There is a strong signal of non-zero circular polarisation at each of the two \ha\ $\sigma$ components, but remarkably, in each of the $\sigma$ components, the polarisation shows both a positive and a negative peak of roughly equal amplitude. The result is that when the mean longitudinal field \bz\ is evaluated by determining the separation between the mean line position as measured in right and left circularly polarised light \citep{Math89,Donaetal97}, the result is $\bz = 0.03 \pm 0.08$\,MG. We could not detect a useful polarisation signal around H$\beta$.

The variation of the $V/I$ signal in each of the $\sigma$ components of H$\alpha$, with clear sign reversal approximately in the middle of the component, suggests that roughly comparable field regions of opposite sign are present on the visible hemisphere at the time of observation, with regions of one sign having local field strength $\vert {\bf B} \vert$ with values perhaps 0.4--0.6\,MG larger than the typical local field strength in the visible region of the opposite polarity. This leads to the polarisation signatures of the two regions occurring at slightly different wavelength separations from the central $\pi$ component, so they do not directly cancel one other. The observed polarisation profiles suggest that if the field is roughly dipolar in global structure, the line of sight during our observation was nearly in the plane of the magnetic equator. 

The lower {\it S/N} spectrum, obtained about 5\,d earlier than the better spectrum, is not significantly different from the better spectrum. There is no clear indication of variability on a time-scale of a few days. 

In addition to the polarisation around H$\alpha$, in both spectra we have detected a weak signal of polarisation in the continuum. The $V/I$ spectrum obtained on 2020-01-08 is offset by $\sim +0.25$\,\%, and the one obtained on 2020-01-13 is offset by $-0.1$\,\% for $\lambda \ga 4500$\,\AA, while at the bluest observed spectral region goes down to $\sim -1$\,\%. For reasons that will be discussed in Sect.~\ref{Sect_Policont_After}, we consider these continuum effects to be probably spurious. 

\subsection{DA star WD\,1703$-$267 = Gaia DR2 4108828945319007744}\label{Sect_WDSeventeen}

%%%%%%%%%%%%%%%%%%%%%%%%%%%%%%%%%%%%%%%%%%%%%%%%%%%%%%%%%%%%%%%%%%%%%%%%%%%%%%%%%%%%%%%%%%%%%%%%%%%%%%%%%%%%%%%%%%%%%%%%%%%
   \begin{figure}[ht]
   \centering
   \includegraphics[width=9cm,trim={0.9cm 2.3cm 0.7cm 1.0cm},clip]{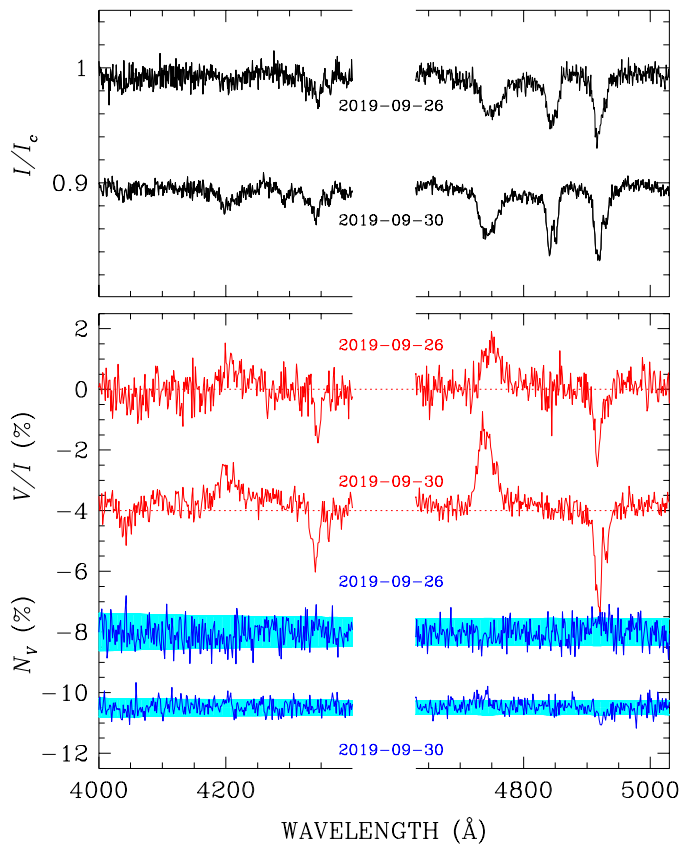}
      \caption{WD\,1703$-$267: Observed $I/I_{\rm c}$ (top panel),  $V/I$ (red curve) and $N/I$ (blue curve) spectra of WD\,1703$-$267 around H$\gamma$ and H$\beta$.
              }
         \label{Fig_wd1703}
   \end{figure}
%%%%%%%%%%%%%%%%%%%%%%%%%%%%%%%%%%%%%%%%%%%%%%%%%%%%%%%%%%%%%%%%%%%%%%%%%%%%%%%%%%%%%%%%%%%%%%%%%%%%%%%%%%%%%%%%%%%%%%%%%%%

This star is a newly identified WD in the \tpv\ \citep{Holletal18}. Although it is reasonably bright and has a significant proper motion of about 100\,mas/yr, it does not seem to have been picked up in proper motion surveys, probably because it is located in a very crowded field only 8$^\circ$ out of the Galactic Plane in the direction of the Galactic Centre. Because the only available name for this object is awkwardly long and not readily recognisable, and because we have consistently found the widely used Villanova WD numbers to be very helpful in identifying individual stars across a range of journal papers, we assign an (unofficial) Villanova WD number to this object. The physical parameters  of \te, $\log g$ and mass for this WD have been estimated by \citet{Holletal18}  and by \citet{Gentetal19}. Both of these studies have fit Gaia three-band photometry and accurate parallaxes to model photometric data and theoretical mass-radius relations to create very similar general calibrations. In addition, \citet{Gentetal19} have improved their calibrations by incorporating SDSS photometry and spectra when these were available. Using these calibrations,  the parameters \te, $\log g$ and mass have been obtained by both these studies for many WDs for which only Gaia data are available. 

The age of WD\,1703--267 has been obtained by interpolation in the on-line cooling curves made available by the Montreal group\footnote{http://www.astro.umontreal.ca/\~{}bergeron/CoolingModels }, which are based on results described by \citet{Tremetal11}. (Ages derived by us from interpolation of the Montreal tables are given with only two digits of precision.) No previous spectroscopy of this object seems to be available. 
   
We have obtained two blue FORS spectra of this star. Both reveal clear Zeeman split H$\beta$ lines with exactly the expected signature of a significant magnetic field in the corresponding $V/I$ spectrum (Fig.\,\ref{Fig_wd1703}). H$\gamma$ is visible in the better spectrum, weakly in an intensity absorption feature at about 4340\,\AA\ and more strongly in polarisation between 4190 and 4400\,\AA. At the low \te\ of this star, the Balmer decrement is quite strong, and H$\delta$ is detectable only in weak polarisation features. Overall, the splitting and polarisation in our spectra are certainly clear enough to establish that this star is a MWD. 
   
\begin{figure}[ht]
   \centering
   \includegraphics[width=9cm,trim={0.0cm 0.0cm 0.7cm 1.0cm},clip]{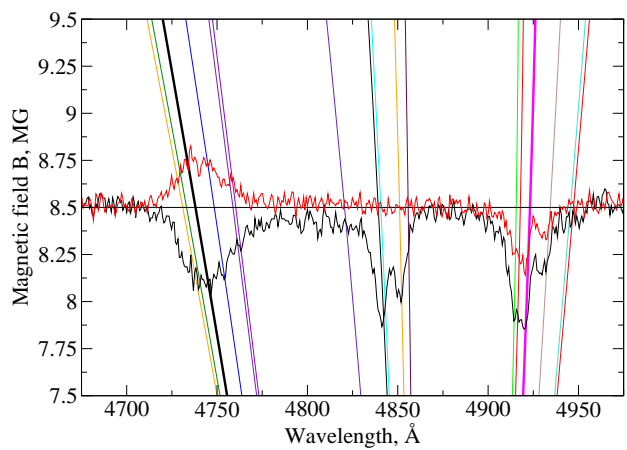}
      \caption{\label{Fig_SW14_WD1703-267} 
      WD\,1703-267: Section of a diagram showing wavelengths magnetically split line components as a function of magnetic field strength $B$ around 8.5\,MG, with 2019-09-30 $I$ (black) and $V$ (red) spectra of WD\,1703--267 superposed. Matching stellar spectra to split line components shows that field producing this spectrum has typical strength of $\bs \approx 8.5$\,MG. We note that wavelengths of blue $\sigma$ components vary more rapidly with $B$ than red components do. 
      }
   \end{figure}

The mean field modulus \bs\ can be estimated from the positions of the two $\sigma$ components of H$\beta$. Both are spread over a few tens of \AA. The mean wavelengths of these components as estimated in both intensity and in circular polarisation, provide us with estimates of \bs. Superimposing the $I$ and $V$ spectra from our two observations shows clearly that the two spectra are significantly different, and yield estimates of $\bs \approx 8.0 \pm 0.5$ and $8.5 \pm 0.5$ for the two observations \citep{SchiWunn14}. One such fit to the (better) spectrum from 2019-09-30 is shown in Fig.\,\ref{Fig_SW14_WD1703-267}, in which we plot the $I$ and $V$ spectra around H$\beta$ on a plot displaying the varying wavelengths of the 18 H$\beta$ line Zeeman components as a function of field strength. The two spectra are positioned vertically to have their continua at about the level corresponding to the deduced magnetic field \bs\ of the star. 

%%%%%%%%%%%%%%%%%%%%%%%%%%%%%%%%%%%%%%%%%%%%%%%%%%%%%%%%%%%%%%%%%%%%%%%
%%%%%%%%%%%%%%%%%%%%%%%%%%%%%%%%%%%%%%%%%%%%%%%%%%%%%%%%%%%%%%%%%%%%%%%%%%%%%%%%%%%%%%%%%%%%%%%%%%%%%%%%%%%%%%%%%%%%%%%%%%%
   \begin{figure}[ht]
   \centering
   \includegraphics[width=9cm,trim={1.0cm 2.5cm 0.8cm 1.0cm},clip]{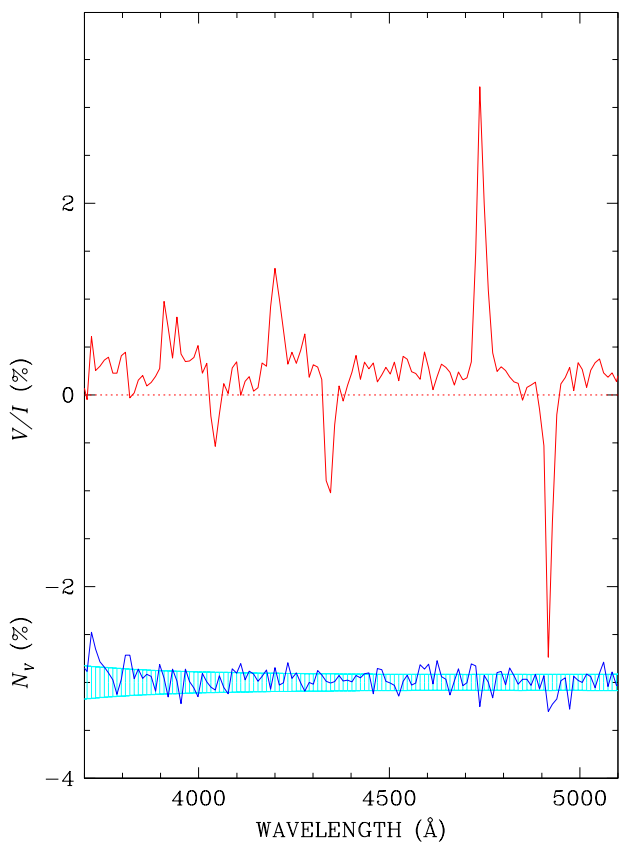}
      \caption{\label{Fig_WD1703_rebinned} 
      WD\,1703$-$267 (observations of September 30, 2019): circular polarisation (red solid line) and null profile (blue solid line, offset by $-3$\,\%), superposed to the $V/I$ error bars (light blue).  Spectra are rebinned at about 11\,\AA. 
      }
   \end{figure}
%%%%%%%%%%%%%%%%%%%%%%%%%%%%%%%%%%%%%%%%%%%%%%%%%%%%%%%%%%%%%%%%%%%%%%%

It is clear from Fig.\,\ref{Fig_SW14_WD1703-267} that the short wavelength $\sigma$ component is broader and shallower than the long wavelength component (about 50\,\AA\ width compared to about 35\,\AA). The Stark broadening should be essentially the same for both components, and at a field strength of about 8.5\,MG the separation  of the strongest subcomponents of the $\sigma$ components is only about 25\,\AA\ and is similar for the two $\sigma$s. However, the variation of the $\sigma$ components with a field strength $B$ is considerably stronger for the blue component than for the red component \citep[see Fig.\,7 of ][]{GarsKemi74}. Hence we consider that the excess broadening of the blue $\sigma$ is due to the spread in field strength over the visible hemisphere. The rather limited excess broadening suggests that the local field strength in the visible hemisphere is mostly within the range of roughly 8.0--9.0\,MG. 
   
The longitudinal field \bz\ is measured, as for the previous case of WD\,1315$-$781, using the wavelength difference between the full H$\beta$ line as seen in right and in left circularly polarised light. In the case of WD\,1703$-$267, there is a substantial separation leading to values of $\bz \sim 2$\,MG (see Table~\ref{Tab_journal}). The small variations in the shape and positions of the Zeeman components of H$\beta$ in the observed flux (Stokes $I$) spectra, and the very distinct differences between the two Stokes $V/I$ spectra, make it clear that, unusually, this MWD is variable on a time scale of a few days. WD\,1703$-$267 is a clear candidate for an observational campaign to obtain enough polarised spectra for modelling the surface magnetic field structure. 
   
The H$\gamma$ line is also clearly detectable in our spectra. In both spectra a weak depression in $I$, centred at about 4340\,\AA, is due mostly to strong, nearly stationary components of the red $\sigma$ component of the line. The effect of the line is much more strongly evident in the $V/I$ spectrum, where a relatively sharp negative excursion in the region 4330--4350\,\AA, and a corresponding broad positive feature between 4190 and 4225\,\AA\ are due to the strongest individual transitions of the red and blue $\sigma$ components of H$\gamma$, leading to polarisation structure very similar to that seen in H$\beta$ \citep{SchiWunn14}. 

We note that the $\pi$ component of H$\beta$ in our second spectrum of WD\,1703$-$267 shows a notch, reminiscent of the line core shape found in double degenerate WD binary systems. It is certainly of interest to consider whether this object is in fact a spectroscopic binary. One possibility is that one of the two minima represents a MWD and the other represents a non-magnetic WD. We can rule this out immediately, as the blue shift of even the redward $\pi$ component minimum is about 10\,\AA\ to the blue of the rest wavelength of H$\beta$ (to which should be added about 1\,\AA\ of gravitational red shift). This would require that the non-magnetic WD is moving away from the MWD at a velocity of the order of 600\,\kms. Therefore both minima must instead be produced in very similar magnetic fields. Most probably, this notch represents an effect of the detailed distribution of the magnetic field over the visible hemisphere of the MWD as viewed during our second observation. 
  
   It is very interesting to consider our data for WD\,1703$-$267 in the light of our earlier effort (see Sec.~2.1) to calibrate the order-of-magnitude relationship of Eq.~(\ref{Eq_Kemp}). For this star, we have accurate values of both \bs\ and \bz, and judging from the simple $V/I$ profiles through the Balmer lines, a relatively simple magnetic field structure. With $\bz \approx 2.75$\,MG in our second observation, we predict from our calibration of the coefficient of Eq.~~(\ref{Eq_Land}) that there should be continuum circular polarisation of $V/I \approx 0.2$\,\%. Rebinning our data over 16 pixels to improve the S/N, we find that a weak continuum circular polarisation is indeed present at the level of $V/I \approx 0.25$\,\% (see Fig~\ref{Fig_WD1703_rebinned}). Furthermore, the relationship between the sign of the continuum polarisation and that of $V/I$ in the blue $\sigma$ components of the Balmer lines is the same as in G99-47. This result reinforces confidence in the use of broad-band circular polarisation as a tool for measuring \bz, and in the general accuracy of our calibration. 

\subsection{DA star WD\,0810$-$353 = UPM\,J0812$-$3529}
This star was identified as a WD within the \tpv\ by the USNO parallax programme \citep{Fincetal18}, and is included in the list of WDs within the \tpv\ established by \citet{Holletal18} on the basis of the Data Release 2 dataset (Gaia Collaboration \citeyear{Gaia18}). For the reasons discussed in connection with WD\,1703$-$267, we provide an (unofficial) Villanova WD number for the star, which we use throughout. Physical parameters based on Gaia astrometry and photometry \citep{Holletal18,Gentetal19} are provided in Table~\ref{Tab_new_mwds}. The spectral class is unknown, as no previous spectra of the star appear to have been taken. As the effective temperature is high enough that we would expect H$\beta$ to be observable if the atmosphere of this object is H-rich, while possible metal lines around 4000\,\AA\ would be of great interest, we obtained circular spectropolarimetry of this star with the 600B and 1200B grisms. WD\,0810$-$353 is bright enough that we were able to obtain quite high {\it S/N} ($450 - 500$ per \AA).

The three Stokes $V/I$ spectra are in quite good agreement in that they all show a region of weak positive polarisation around 4720\,\AA, and two regions of stronger negative polarisation around 4830 and 4960\,\AA. 

\begin{figure}[ht]
   \centering
   \includegraphics[width=9cm,trim={0.9cm 2.5cm 0.7cm 1.0cm},clip]{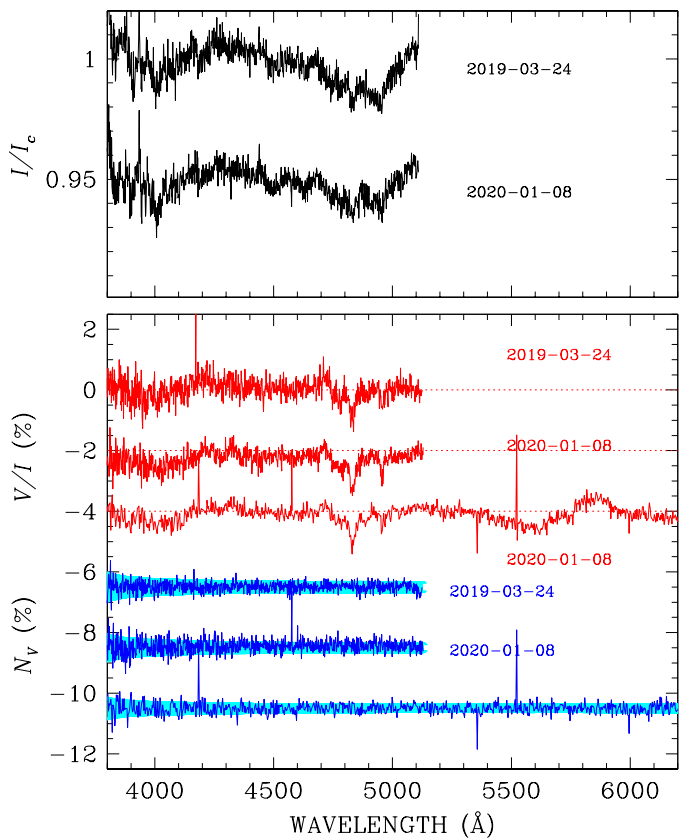}
      \caption{WD\,0810$-$353: observed flux normalised to the continuum as explained in the text (top panel, black solid lines with error bars); $V/I$ (bottom panel red solid lines) and null profiles $N_V$ (bottom panel, blue solid line), offset for display purpose, and superposed to the $V/I$ error bars (in light blue). 
      }
         \label{Fig_wd0810}
   \end{figure}

The $I$ spectra obtained with grism 1200B show a series of weak depressions (with an amplitude of the order of 1\,\%) that are hard to disentangle from instrumental features, even after applying flat-fielding correction. After an approximate normalisation to the continuum (clearly difficult to identify), the ripple in Stokes $I$ show some resemblance to the shape of the $V$ spectra. In an attempt to further reduce instrumental features in Stokes~$I$, in Fig.~\ref{Fig_wd0810} we show the ratio between the flux of WD\,0810$-$353 and that of an unpolarised DC star, WD\,2017$-$306, that was observed with the same grism and same CCD, normalised to the continuum. We note that the weak dips in the flux correspond to the Stokes $V$ features at approximately 4780, 4830 and 4956\,\AA. Since circular polarisation measurements are virtually free from flat-fielding issues \citep[e.g.][]{Bagetal09}, the fact that Stokes $V$ features coincide with Stokes $I$ features suggest that also the latter ones are real and not due to instrument features.

The fact that line-like features are observed in Stokes~$V$, and are correlated with structure in $I$, strongly suggests that this star has an H-rich atmosphere, as a He-rich atmosphere would not lead to any line-like features in the visible window. Thus we try to estimate the field on the assumption that this star is a DA. A first possible interpretation of the polarised spectral features of WD\,0810$-$353 between 4700\,\AA\ and 5000\,\AA\ is that they are due to a very weak H$\beta$ line, split and mildly shifted by a magnetic field of a few MG. Because this star has a value of \te\ very similar to that of WD\,1703$-$267 (Sect.~\ref{Sect_WDSeventeen}), a field as weak as a few MG should produce a fairly strong but obviously Zeeman-split H$\beta$ rather like the one observed in WD\,0810$-$353, with a similar form for the $V/I$ profile. A field of this strength could account for the depressions in $I$ (although not for their weakness), and for the local polarisation peaks of opposite sign at 4720 and 4960\,\AA. However, this interpretation does not provide any explanation for the stronger polarisation feature in between these two peaks, near 4830\,\AA. If this were due to a weaker field than 10\,MG on part of the visible hemisphere, there should be a corresponding polarisation feature of positive sign a little redward of the unshifted position of H$\beta$, but none is visible. Thus the interpretation of these features as being due to mild Zeeman splitting of H$\beta$ is not convincing. 

We next consider the possibility that the field is considerably stronger than $\sim 10$\,MG, say in the range of 20--50\,MG. This situation would lead to the 18 strong individual components of H$\beta$ being spread more or less uniformly over much of the region from about 4000 to 5100\,\AA. Both the division of the oscillator strength of H$\beta$ (already quite weak because of the low \te\ of the MWD) between many components, and the very probable variation of field strength (and thus of component absorption wavelengths) over the visible stellar hemisphere, would easily account for the extreme weakness and lack of obvious pattern of the features in $I$. Instead, we would need to look primarily to the $V/I$ spectrum for hints about the field strength. 

\begin{figure}[ht]
   \centering
   \includegraphics[width=9cm,trim={0.0cm 0.0cm 0.7cm 1.0cm},clip]{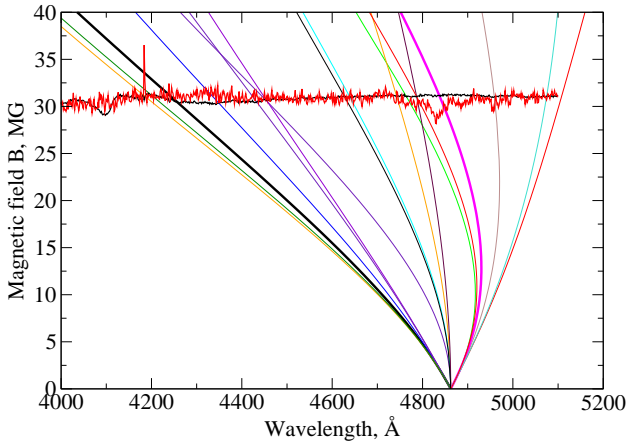}
      \caption{WD\,0810$-$353: Flux (black) and polarisation (red) spectra superposed on a magnetic splitting diagram. The notch in $V$ coincides with the strongest red \hb\ component (heavy magenta curve) at about 30\,MG. 
      }
         \label{Fig_SW14-Hb-wd0810-353}
   \end{figure}

In the $V/I$ spectrum, the most prominent features are the two negative dips at 4830 and 4956\,\AA. Because the feature at 4956\,\AA\ is particularly narrow, we look for a strong $\sigma$ component of H$\beta$ that is roughly stationary with changes in field strength at about this wavelength. The $\sigma$ component due to the 2p0--4d-1 transition has this behaviour, and is centred on the correct wavelength for field strength of about 15 or about 30\,MG. If we then try to associate the bluer negative $V/I$ feature with another, somewhat more rapidly varying $\sigma$ component, the very strong 2p-1--4d-2 transition goes through the correct wavelength range at about 30\,MG; see Fig.\,\ref{Fig_SW14-Hb-wd0810-353}.

Our 600B spectrum covers the additional spectral range between 5130 and 6230\,\AA. In this spectral window there are no obvious flux spectral features. In $V/I$, however, there is a clear polarisation "S-wave" extending between about 5400 and 6000\,\AA. The interpretation of this feature is not obvious, although the long-wavelength part of the S-wave may be due to rapidly changing blue $\sigma$ components of \ha, which reach this wavelength window in fields of about 30\,MG or more. In any case, the presence of this non-zero, structured broad continuum circular polarisation certainly confirms the basic conclusion that WD\,0810$-$353 is a MWD. 

In summary, our tentative conclusion is that WD\,0810$-$353 is a DA MWD whose weak Balmer lines have been reduced almost to invisibility in the $I$ spectrum by the strong magnetic splitting into many components produced by a field of the order of $\bs \sim 30$\,MG. This conclusion is based mainly on the presence of a small number of clear features in the $V/I$ spectrum. We do not have a useful estimate of \bz, except that the presence of significant narrow and broad band circular polarisation strongly suggests that it is non-zero. Further observations around \ha\ would be very helpful in testing this interpretation of the observations.

%%%%%%%%%%%%%%%%%%%%%%%%%%%%%%%%%%%%%%%%%%%%%%%%%%%%%%%%%%%%%%%%%%%%%%%%%%%%%%%%%%%%%%%%%%%%%%%%%%%%%%%%%%%%%%%%%%%%%%%%%%%
%%%%%%%%%%%%%%%%%%%%%%%%%%%%%%%%%%%%%%%%%%%%%%%%%%%%%%%%%%%%%%%%%%%%%%%%%%%%%%%%%%%%%%%%%%%%%%%%%%%%%%%%%%%%%%%%%%%%%%%%%%%

%%%%%%%%%%%%%%%%%%%%%%%%%%%%%%%%%%%%%%%%%%%%%%%%%%%%%%%%%%%%%%%%%%%%%%%%%%%%%%%%%%%%%%%%%%%%%%%%%%%%%%%%%%%%%%%%%%%%%%%%%%%
%%%%%%%%%%%%%%%%%%%%%%%%%%%%%%%%%%%%%%%%%%%%%%%%%%%%%%%%%%%%%%%%%%%%%%%%%%%%%%%%%%%%%%%%%%%%%%%%%%%%%%%%%%%%%%%%%%%%%%%%%%%

\subsection{DC star WD\,0004$+$122 = LP464-57} 

%%%%%%%%%%%%%%%%%%%%%%%%%%%%%%%%%%%%%%%%%%%%%%%%%%%%%%%%%%%%%%%%%%%%%%%%%%%%%%%%%%%%%%%%%%%%%%%%%%%%%%%%%%%%%%%%%%%%%%%%%%%
   \begin{figure}[ht]
   \centering
   \includegraphics[width=9cm,trim={0.9cm 2.5cm 0.7cm 9.8cm},clip]{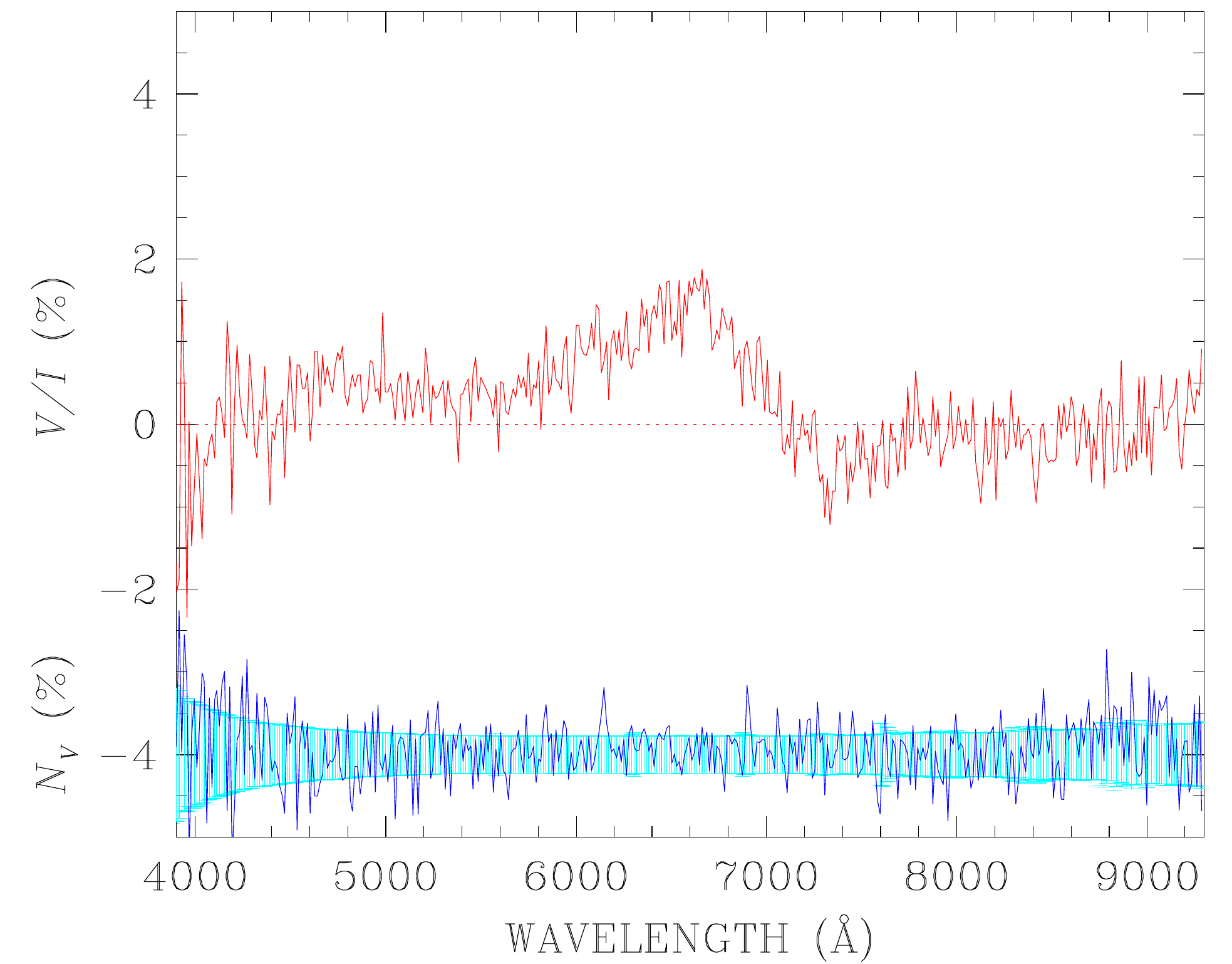}
      \caption{WD\,0004$+$122: $V/I$ (red solid line) and null profiles $N_V$ (bottom panel, blue solid line), offset by -4.0\,\% for display purpose, and superposed on the $V/I$ error bars). 
              }
         \label{Fig_wd0004}
   \end{figure}
%%%%%%%%%%%%%%%%%%%%%%%%%%%%%%%%%%%%%%%%%%%%%%%%%%%%%%%%%%%%%%%%%%%%%%%%%%%%%%%%%%%%%%%%%%%%%%%%%%%%%%%%%%%%%%%%%%%%%%%%%%%

%%%%%%%%%%%%%%%%%%%%%%%%%%%%%%%%%%%%%%%%%%%%%%%%%%%%%%%%%%%%%%%%%%%%%%%%%%%%%%%%%%%%%%%%%%%%%%%%%%%%%%%%%%%%%%%%%%%%%%%%%%%
   \begin{figure}[ht]
   \centering
   \includegraphics[width=9cm,trim={0cm 0.0cm 0.7cm 2.0cm},clip]{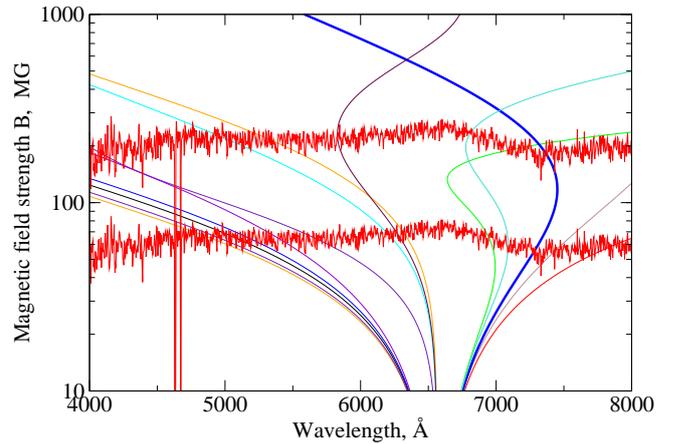}
      \caption{WD\,0004+122: $V/I$ spectrum of WD\,0004+122 (red) superposed twice on diagram showing wavelengths of \ha\ components as a function of field strength. Spectra are placed at approximate levels where 7330\,\AA\ polarisation feature coincides with strongest red $\sigma$ transition 2p-1--3d-2 (heavy blue curve), showing possible interpretation of feature as produced by either $\bs \approx 60$ or $\bs \approx 200$\,MG field.). 
              }
         \label{Fig_wd0004_sw14}
   \end{figure}
%%%%%%%%%%%%%%%%%%%%%%%%%%%%%%%%%%%%%%%%%%%%%%%%%%%%%%%%%%%%%%%%%%%%%%%%%%%%%%%%%%%%%%%%%%%%%%%%%%%%%%%%%%%%%%%%%%%%%%%%%%%

This star of high proper motion was first noted in the Luyten--Palomar survey \citep{SalGou03}. The star (the faintest of our sample) was classified as a DC WD by \citet{KawkVenn06}, who estimated $\te = 6300$\,K. It has been studied by several groups. Recent modelling \citep{Limoetal15,Blouetal19} has reduced the value of \te\ to about 5000\,K, and provided values of $\log g$, mass and age. \citet{Holletal18} have identified this WD as lying within the \tpv. Astrometric and physical parameters of WD\,0004$+$122 are listed in Table\,\ref{Tab_new_mwds}. Its estimated mass is very close to the average mass for all local WDs. We obtained one observation of this star with grism 300V.

The flux spectrum $I$ shows no stellar absorption features. The circular polarisation spectrum  $V/I$ is displayed in Fig.~\ref{Fig_wd0004} with a red solid line.  It shows clearly non-zero values over most of the recorded spectral range. One probably real feature is a narrow negative polarisation dip at about 7330\,\AA. 

It is helpful to compare the $V/I$ spectrum of this star to two other cool MWDs for which non-zero circular polarisation is detected and for which reasonably high-resolution polarisation spectra extending to 8000\,\AA\ are available: WD\,1829+547 = G\,227-35 \citep{Coheetal93,PutnJord95}, which has $\te \approx 6330$\,K, and WD\,0756+437 = G\,111-49 \citep{Putn95}, with $\te \approx 7000$\,K. In both of these stars the strongest narrow polarisation feature in the spectrum is a line-like feature near 7300--7400\,\AA. \citet{Coheetal93} have argued that this feature is due to the stationary 2p-1--3d-2 component of \ha\ in a field of approximately 120\,MG, so that both these stars have H-rich atmospheres.\footnote{To make this comparison we recall that because of the difference in the definition of the sign of $V$, the feature in question appears as a sharp dip in the $V/I$ spectrum of WD\,0004$+$122 of Fig.~\ref{Fig_wd0004}, while it appears as a spike in the $V/I$ spectra of WD\,1829+547 presented by \citet{Coheetal93} and \citet{PutnJord95} and of WD\,0756+437 presented by \citet{Putn95} .}

It is plausible that this strongest stationary component of \ha\ is also the cause of the weak 7330\,\AA\ polarisation feature detected in WD\,0004$+$122. This \ha\ component would be expected to be extremely weak at the very low \te\ value of this MWD, but it might still be detectable in polarisation even if no longer in flux. Thus we tentatively conclude that WD\,0004$+$122 has a H-rich atmosphere. 

Because the wavelength of this feature is a little shorter than that of the turning point of the relevant stationary line, as illustrated in Fig.\,\ref{Fig_wd0004_sw14} \citep{SchiWunn14},  we estimate the field strength to be either $\bs \sim 60$ or  $\bs \sim 200$\,MG. Using the approximate calibration of Eq.\,(\ref{Eq_Land}) with the peak continuum polarisation observed, we would estimate perhaps $\bz \sim 30$\,MG. If \bs\ is near 200 MG, the relatively weak continuum polarisation would suggest that the dipole component of the field makes a rather large angle with the line of sight; if \bs\ is near 60 MG the larger ratio of \bz\ to \bs\ suggests a nearly pole-on view of the field. Further observation to confirm the reality of the weak polarisation feature at 7330\,\AA\ would be very helpful. 

It is also of interest to note that a  morphologically very similar polarisation large-scale spectrum is observed in the large field MWD WD\,0756+437 \citep{Putn95}. Like WD\,0004+122, WD\,0756+437 shows polarisation amplitude increasing strongly towards the blue from about 7000\,\AA\ to a peak, and then sharply declining. Both the amplitude of the broad-band shape and the polarisation feature near 7400\AA\ in WD\,0756+437 are much stronger than those of WD\,0004+1222, presumably because of higher \te. Furthermore, as in WD\,0004+122 the sign of the 7400\,\AA\ feature is opposite to that of the continuum hump. These striking similarities provide substantial support for our interpretation of the spectrum of WD\,0004+122.

\subsection{DC star WD\,0708$-$670 = SCR J0708-6706}

%%%%%%%%%%%%%%%%%%%%%%%%%%%%%%%%%%%%%%%%%%%%%%%%%%%%%%%%%%%%%%%%%%%%%%%%%%%%%%%%%%%%%%%%%%%%%%%%%%%%%%%%%%%%%%%%%%%%%%%%%%%
   \begin{figure}[ht]
   \centering
   \includegraphics[width=9cm,trim={0.9cm 2.5cm 0.7cm 9.8cm},clip]{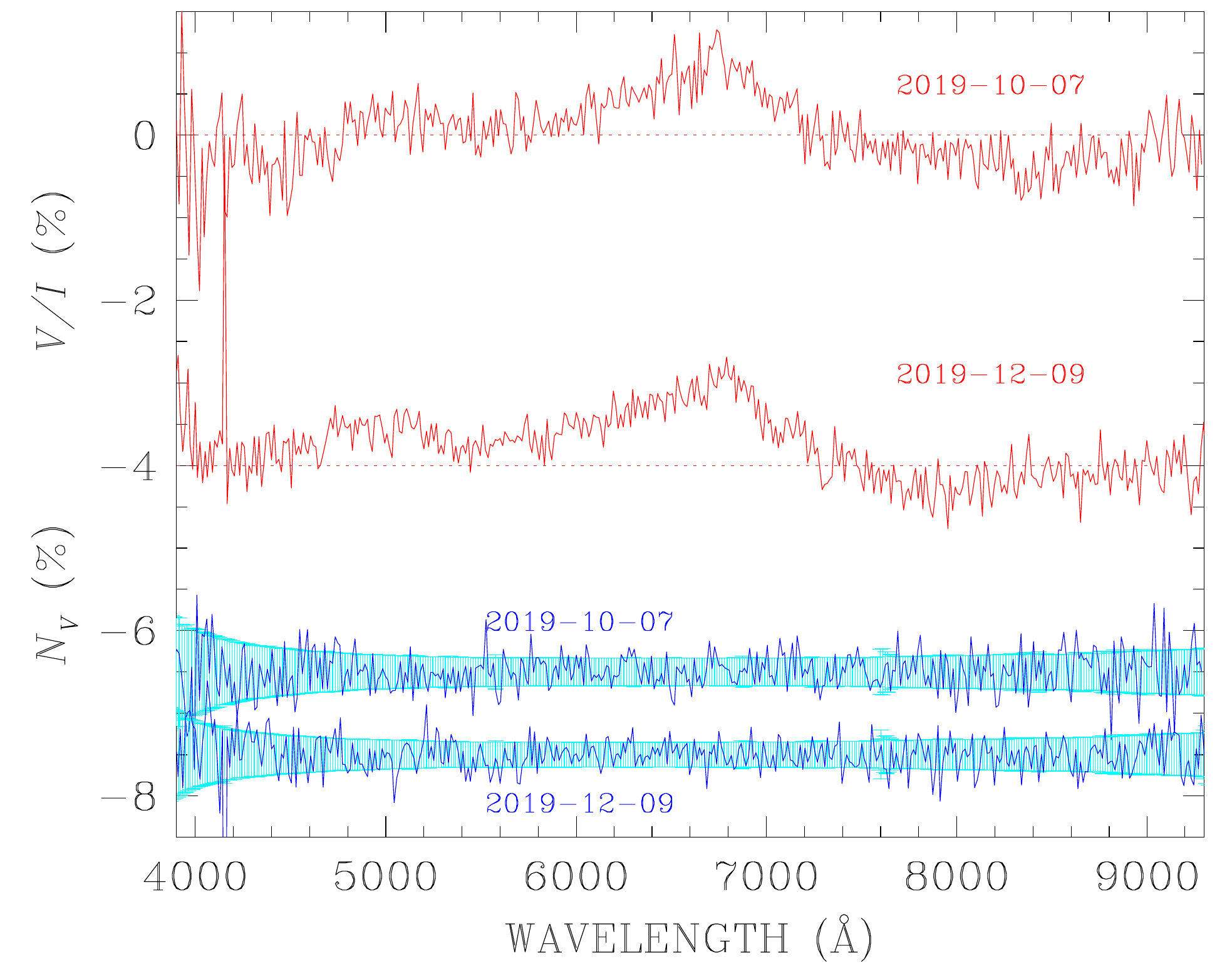}
      \caption{WD\,0708$-$670: Stokes
       $V/I$ (red lines and null profiles (blue solid lines). The $V/I$ spectrum of 2019-12-09 is offset by $-4$\,\% for display purpose, offset by $-7$ and $-8$\,\%, superposed to the $V/I$ error bars. Light dotted lines provide zero levels for the two $V/I$ spectra. }
         \label{Fig_wd0708}
   \end{figure}
%%%%%%%%%%%%%%%%%%%%%%%%%%%%%%%%%%%%%%%%%%%%%%%%%%%%%%%%%%%%%%%%%%%%%%%%%%%%%%%%%%%%%%%%%%%%%%%%%%%%%%%%%%%%%%%%%%%%%%%%%%%

This star was first noticed as a faint high-proper motion star by \citet{Fincetal07}. It has since been identified as a DC WD, without any stellar spectral lines. With an effective temperature of about 5000\,K \citep{Subaetal08,Giametal12,Subaetal17}, it could have an atmosphere dominated by either H or He, as a WD with an H-rich atmosphere may well show at most an extremely weak \ha\ Balmer line at this \te, and of course an He-rich atmosphere would certainly show no lines at all unless it is polluted with H or metals. We therefore obtained two high {\it S/N} circular polarisation spectra with grism 300V from 3500 to 9300\,\AA. Both of the observed $V/I$ spectra are shown in Fig.\,\ref{Fig_wd0708} (one offset by $-4$\,\% for display purpose). As for WD\,0004$+$122, the flux (Stokes $I$) spectra are essentially featureless, while both circular polarisation spectra are clearly non-zero, and show no obvious sign of variability on a time-scale of two months. WD\,0708$-$670 is definitely a MWD. 

The two circular polarisation spectra are not only very similar to each other, but also very similar to that of WD\,0004$+$122, even to the point of apparently showing a still somewhat weaker but apparently real feature in the $V/I$ spectrum at about 7310\,\AA. This feature is quite clear in the second spectrum, and may well be present in the (noisier) first spectrum. Again, this feature is the most prominent narrow structure in the $V/I$ spectrum. 

The overall similarity of the shape and amplitude of the $V/I$ spectra of WD\,0708$-$670 to that of WD\,0004$+$122, and morphologically to that of WD\,0756+437, certainly encourages us to consider applying the same reasoning to this MWD. Again, we conclude that this MWD has also a H-rich atmosphere, with a field of $\bs \sim 200$\,MG or $\sim 60$\, MG, and a rather weak longitudinal field of the order of $\bz \sim 15$\,MG.

\subsection{DC star WD\,2049$-$253 = Gaia DR2 6805792571514600960}

%%%%%%%%%%%%%%%%%%%%%%%%%%%%%%%%%%%%%%%%%%%%%%%%%%%%%%%%%%%%%%%%%%%%%%%%%%%%%%%%%%%%%%%%%%%%%%%%%%%%%%%%%%%%%%%%%%%%%%%%%%%
   \begin{figure}[ht]`
   \centering
   \includegraphics[width=9cm,trim={0.9cm 2.5cm 0.7cm 9.8cm},clip]{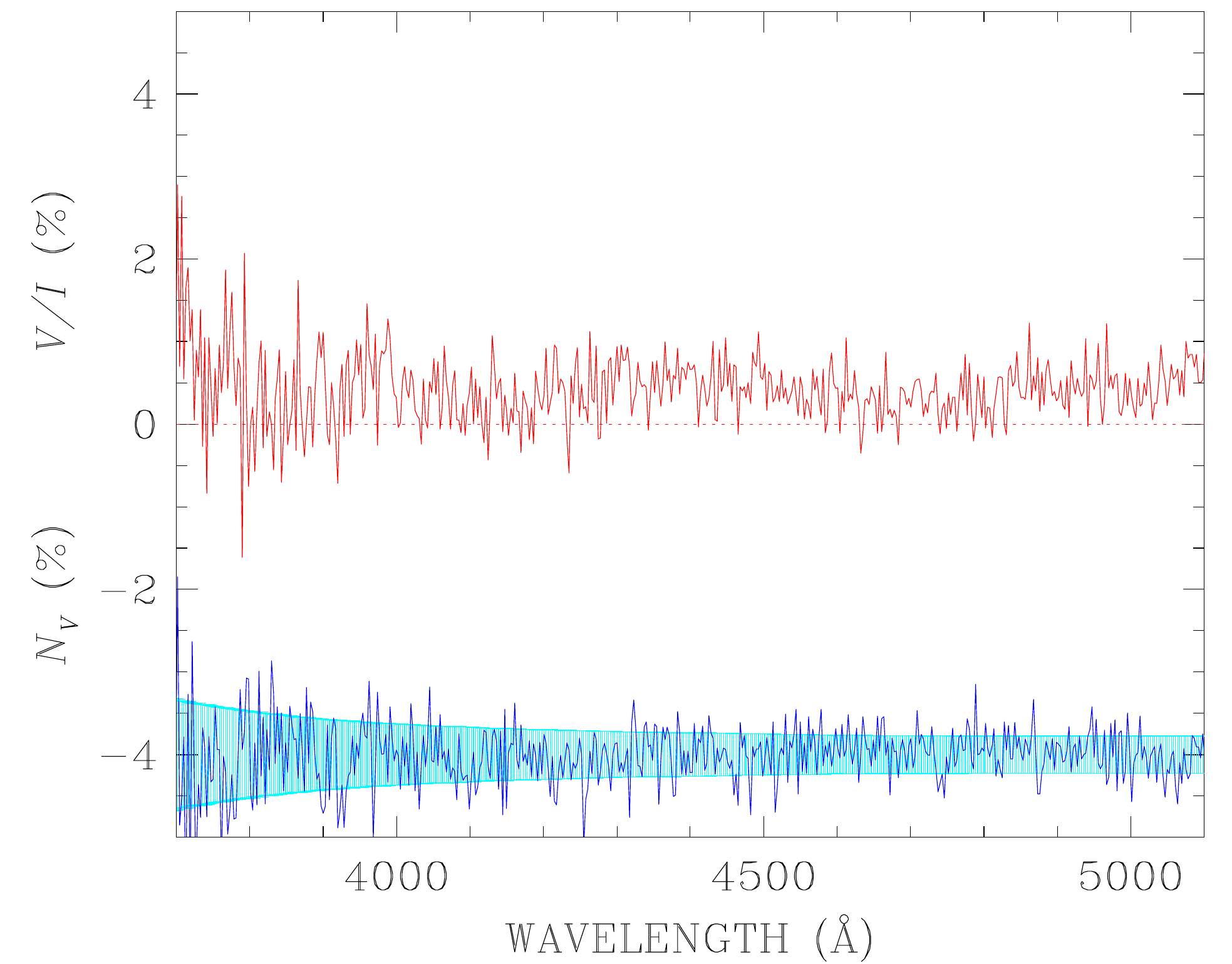}
      \caption{WD\,2049$-$253: Same as Fig.~\ref{Fig_wd0004} for WD\,2049$-$253.} 
         \label{Fig_wd2049}
   \end{figure}
%%%%%%%%%%%%%%%%%%%%%%%%%%%%%%%%%%%%%%%%%%%%%%%%%%%%%%%%%%%%%%%%%%%%%%%%%%%%%%%%%%%%%%%%%%%%%%%%%%%%%%%%%%%%%%%%%%%%%%%%%%%
  
This star was first identified as a WD within the \tpv\ by \citet{Holletal18}. It is a moderately faint object of high proper motion in a sparsely populated southern field, which has probably escaped previous notice by falling below the magnitude limits of proper motion surveys. There is no spectroscopic information about the WD. Basic physical parameters of the star have been estimated by \citet{Holletal18} on the assumption that the atmosphere is dominated by H, and by \citet{Gentetal19} for both H and He-rich atmospheric compositions. 
  
Our spectropolarimetry, obtained with grism 1200B, reveals that WD\,2049$-$253 is a DC star without the slightest hint of spectral features in the blue Stokes $I$ spectrum, so we are unable to determine the atmospheric composition. However, the parameters determined under the two different assumptions about the atmospheric chemistry are at most marginally different. In Table~\ref{Tab_new_mwds} we have listed the properties derived by \citet{Gentetal19} assuming an H-rich atmosphere. 
  
The circular polarisation spectrum is equally devoid of spectral features. However, as seen in Fig.~\ref{Fig_wd2049}, the entire observed circular polarisation spectrum is offset from zero by approximately 0.5\.\%. The null profile, computed from the same data, oscillates around zero well within its error bars. 
  
The roughly constant value of $V/I$ through the blue spectral region resembles the continuum polarisation of WD\,0553+053 = G99--47 discussed in Sect.~\ref{Sect_Estimates}. The observed level of polarisation suggests that the line-of-sight field component visible during the observation was $\bz \approx 7$\,MG, and then by inference that $\bs \geq 20$\,MG.   

Our detection should be confirmed by further observations, that ideally should be carried out with grism 300V to cover a wider spectral range, specifically including the range around \ha.

\section{Comments on the reliability of our detections}\label{Sect_Policont_After}
In this section we discuss the possibility that some of our field detections in the WDs stars may be spurious. 
As a Cassegrain mounted instrument, we do not expect FORS2 to be strongly affected by instrumental polarisation. However, in our survey we are interested to know whether circular polarisation signals as small as a few times $10^{-3}$ are real or spurious, as this is a regime that has rarely been explored with FORS2 in circular spectropolarimetric mode.  The instrument's capabilities in the continuum (in both linear and circular polarisation) have been discussed in Sect.~7 of \citet{Bagetal09}, in Sects.~3.2 and 3.3 of \citet{Sieetal14}, and in Sects. 3.1, 3.2 and 3.3 of \citet{Bagetal17}. The conclusion is that cross-talk from $I$ to $V$ should be $\le 10^{-3}$ \citep[see in particular Fig.~10 of][]{Sieetal14}. Our survey provides us the possibility of confirming that cross-talk from intensity has an amplitude of a few times $10^{-4}$. Most of the DA WDs that we have observed with FORS2 are not magnetic, or, if they are, the magnetic field is of the order of a few kG at most; otherwise, we would detect it from the polarisation of the Balmer lines. Therefore all DA WDs in which we have not detected a magnetic field should also have unpolarised continuua. Even a quick inspection to Fig.~A.1 of \citet{BagnLand18} shows that all the stars displayed (both those observed with FORS and with ISIS) have continuum circular polarisation spectra consistent with zero within the noise. 

We should note, however, that either the collimating lens or the longitudinal atmospheric dispersion corrector act as a weak waveplate in front of the polarising optics, causing a non-negligible cross-talk from linear to circular polarisation. We estimate that up to 10\,\% of the signal of linear polarisation could be transformed into circular polarisation (see Sect.~7.4 of \citeauthor{Bagetal09} \citeyear{Bagetal09}, and in particular their Fig.~10; see also Fig.~9 of \citeauthor{Sieetal14} \citeyear{Sieetal14}).  

The observed WDs are all within the local interstellar bubble, and we do not expect our targets to show any significant signal of linear polarisation due to the interstellar medium \citep{Lero99}. However, it must be noted that contaminating background light may be strongly polarised if the observations are performed with some lunar illumination. In that case, because of the cross-talk from linear to circular polarisation, an imperfect background subtraction may lead to spurious detection of circular polarisation. The risk of inaccurate background subtraction is particularly significant with FORS2 because its cross-talk changes rapidly across the field of view. Background polarisation, when measured at a certain distance from the stellar spectrum, might be not representative of the source background; in fact, cross-talk might even change sign \citep[see Fig.~10 of][]{Bagetal09}.

In order to evaluate the combined effect of sky polarisation, cross-talk, and imperfect background subtraction, it is useful to consider how the stellar polarisation is estimated from the measured quantities. The fraction of polarisation $\pv = V/I$ of the star is obtained after subtracting the sky polarisation using the following relationship:
%%%%%%%%%%%%%%%%%%%%%%%%%%%%%%%%%%%%%%%%%%%%%%%%%%%%%%%%%%%%%%%%%%%%%%%%%%%%%%%%%%%%%%%%%%%%%%%%%%%%%%%%%%%%%%%%%%%%%%%%%%%%%%%%
\begin{equation}
%    \pv^{(*)} = \frac{\pv^{\rm (obs)}\,I^{\rm (obs)} - \pv^{\rm (bkg)}\,I^{\rm (bkg)}}{I^{\rm (obs)} - I^{\rm (bkg)}} = X - Y 
    \pv^{(*)} = X - Y 
\label{Eq_Pol}
\end{equation}
%%%%%%%%%%%%%%%%%%%%%%%%%%%%%%%%%%%%%%%%%%%%%%%%%%%%%%%%%%%%%%%%%%%%%%%%%%%%%%%%%%%%%%%%%%%%%%%%%%%%%%%%%%%%%%%%%%%%%%%%%%%%%%%%
where 
%%%%%%%%%%%%%%%%%%%%%%%%%%%%%%%%%%%%%%%%%%%%%%%%%%%%%%%%%%%%%%%%%%%%%%%%%%%%%%%%%%%%%%%%%%%%%%%%%%%%%%%%%%%%%%%%%%%%%%%%%%%%%%%%
\begin{equation}
    X = \frac{\pv^{\rm (obs)}\,I^{\rm (obs)}}{I^{\rm (obs)} - I^{\rm (bkg)}}\ \ \ \ Y=\frac{\pv^{\rm (bkg)}\,I^{\rm (bkg)}}{I^{\rm (obs)} - I^{\rm (bkg)}} \; ,
\label{Eq_Def}
\end{equation} 
%%%%%%%%%%%%%%%%%%%%%%%%%%%%%%%%%%%%%%%%%%%%%%%%%%%%%%%%%%%%%%%%%%%%%%%%%%%%%%%%%%%%%%%%%%%%%%%%%%%%%%%%%%%%%%%%%%%%%%%%%%%%%%%%
and the indices $(*)$, (bkg) and (obs) refer to the star, the background, and on the quantites measured prior to background subtraction, respectively. These equations, which will be used below, help to understand why a small inaccuracy in the estimate of the background polarisation may have a non-negligible impact in the measurement of the polarisation of faint stars. We note that the WDs observed by \citet{BagnLand18} are all relatively bright (most of them have $V \la 14$, and many $V \la 13$); however, some of those presented in this paper have $V \ga 16$, and this is why they will require a more detailed discussion on individual stars. We finally note that, while cross-talk from linear to circular polarisation is a severe problem for the observations with the FORS2 instrument, the ISIS instrument of the ING WHT (which was used for other observations of our survey not presented in this paper), is free from this effect \citep{BagnLand19a}.

\subsection{WD\,0004$+$122 and WD\,0708$-$670}
Stars WD\,0004$+$122 and WD\,0708$-$670 are both of spectral class DC, and are among the faintest ones observed in our survey; these cases may look suspicious because these stars exhibit very similar polarisation spectra; so far we have never seen two WDs with continuum polarised spectra that look so similar to each other. Therefore, it is possible that their polarisation may be due to an instrumental signature.

\begin{figure}
   \centering
   \includegraphics[width=9cm,trim={0.9cm 6.5cm 0.7cm 3.0cm},clip]{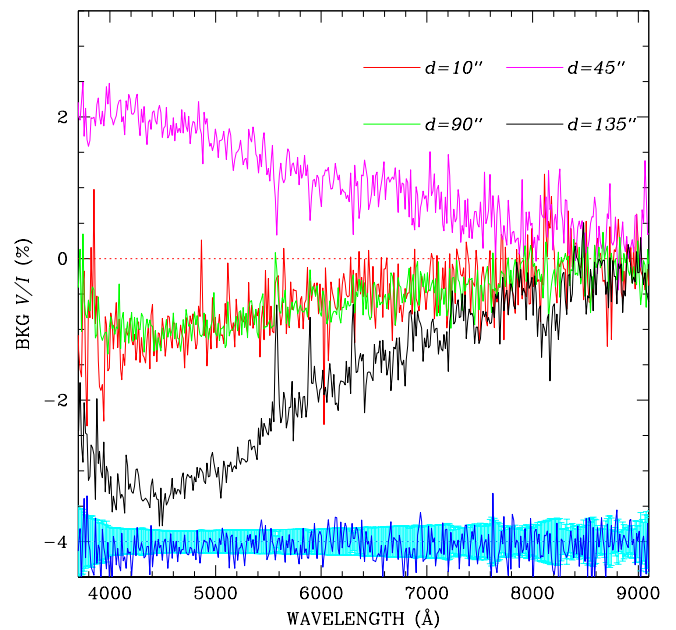}
      \caption{Fraction of circular polarisation of the background around star WD\,0708$-$670
      observed on 2019-10-17. The red, magenta, green and black solid lines show {\it V/I} at various distances from the star as indicated in the Figure. The blue solid line represents the null profile (offset by $-4$\% for display purpose, and with errorbars in light blue colour) corresponding to the {\it V/I} spectrum measured at 45\arcsec\ from the star (magenta line). This profile is representative of all other null profiles.} 
         \label{Fig_BKG_WD0708}
   \end{figure}

\begin{figure}
   \centering
   \includegraphics[width=9cm,trim={2.9cm 6.5cm 1.7cm 3.0cm},clip]{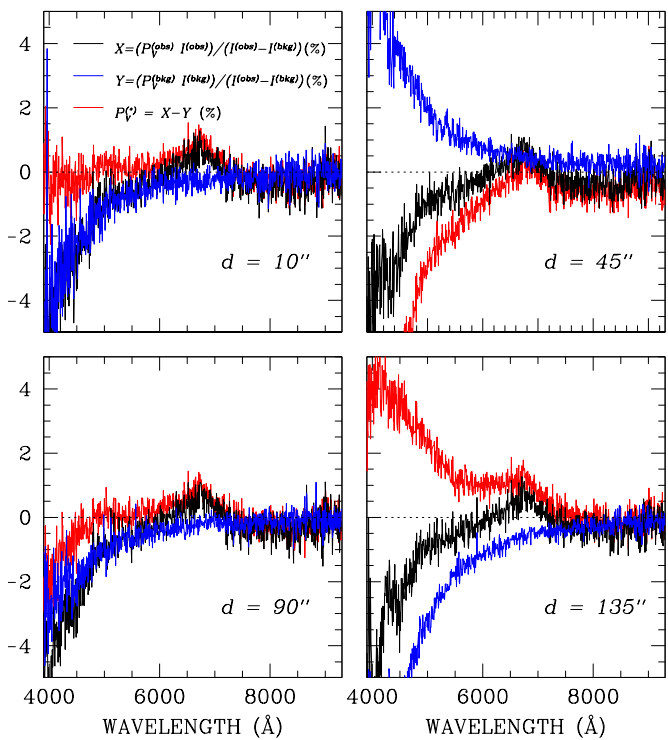}
      \caption{Observations of WD\,0708$-$670 obtained on 2019-10-17 in the presence of strong
      lunar background. The star's polarisation (red solid lines) is obtained after subtracting the background contribution (blue solid lines) estimated from different regions of the CCD from the total observed polarisation (black solid lines). In each panel, the value of $d$ refers to the distance from the source of the region where the background was estimated. See text for more details. } 
         \label{Fig_BKG_Sub}
   \end{figure}

On night October 17, 2019, star WD\,0708$-$670  was observed at $\sim 70\degr$\ from the Moon, when the fraction of lunar illumination was 0.65. Therefore we do expect a high linearly polarised background that is converted into circular polarisation in a way that depends on the position in the instrument field of view. Figure~\ref{Fig_BKG_WD0708} shows the polarisation of the background measured at various angular distances $d$ from the target, and fully confirms our predictions. With reference of Fig.~2.7 of the FORS2 instrument manual, we have measured the background in the area immediately surrounding the star under slit No.~10 ($d=10\arcsec$); in the area under slit No.~8 ($d=45\arcsec$); in the area under slit No.~6 ($d=90\arcsec$); and in the area under slit No.~4 ($d=135\arcsec$).
%%%%%%%%%%%%%%%%%%%%%%%%%%%%%%%%%%%%%%%%%%%%%%%%%%%%%%%%%%%%%%%%%%%%%%%%%%%%%%%%%%%%
%%%%%%%%%%%%%%%%%%%%%%%%%%%%%%%%%%%%%%%%%%%%%%%%%%%%%%%%%%%%%%%%%%%%%%%%%%%%%%%%%%%%%
To properly evaluate the potential impact of sky polarisation it is necessary to weight it by the sky intensity relative to the star intensity. This is done in
Fig.~\ref{Fig_BKG_Sub}, which shows the quantities $X$ and $Y$ defined in Eq.~(\ref{Eq_Def}), and their difference, which gives the star's polarisation (see Eq.~\ref{Eq_Pol}). It appears that the measured star's polarised spectrum has a different shape according to the CCD area selected to estimate the background.
%%%%%%%%%%%%%%%%%%%%%%%%%%%%%%%%%%%%%%%%%%%%%%%%%%%%%%%%%%%%%%%%%%%%%%%%%%%%%%%%%%%%
%%%%%%%%%%%%%%%%%%%%%%%%%%%%%%%%%%%%%%%%%%%%%%%%%%%%%%%%%%%%%%%%%%%%%%%%%%%%%%%%%%%%%
%%%%%%%%%%%%%%%%%%%%%%%%%%%%%%%%%%%%%%%%%%%%%%%%%%%%%%%%%%%%%%%%%%%%%%%%%%%%%%%%%%%%
   \begin{figure}
   \centering
   \includegraphics[width=9cm,trim={2.9cm 6.5cm 1.7cm 3.0cm},clip]{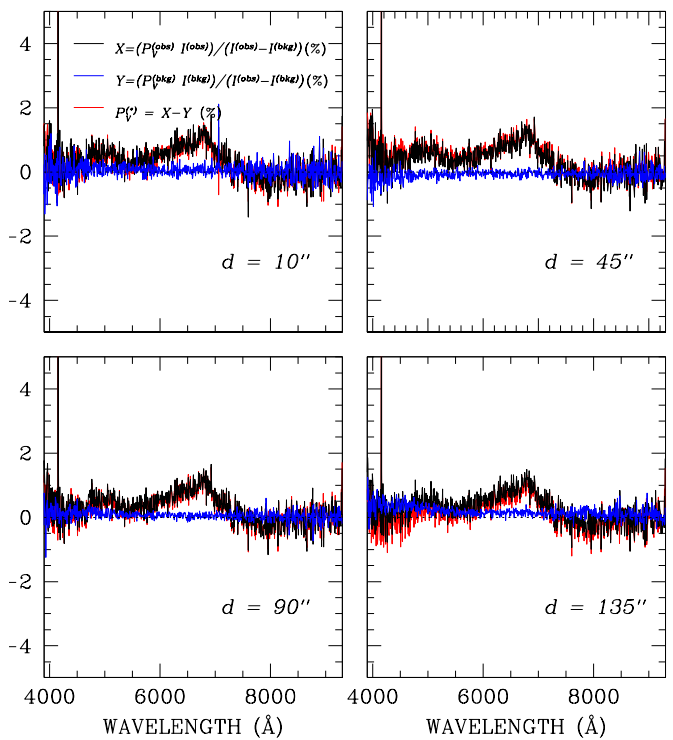}
      \caption{Same as Fig.~\ref{Fig_BKG_Sub}, but for the observations of WD\,0708$-$670 obtained on 2019-12-09 with almost no lunar background.} 
         \label{Fig_noBKG_WD0708}
   \end{figure}
%%%%%%%%%%%%%%%%%%%%%%%%%%%%%%%%%%%%%%%%%%%%%%%%%%%%%%%%%%%%%%%%%%%%%%%%%%%%%%%%%%%%%

A particularly interesting comparison is that between the plots of the quantities $X$, $Y$, and $V/I$ with background calculated from different regions of the CCDs, and the same quantities for the same star calculated from the observations of night December 9 2019, when the moon was setting. Figure~\ref{Fig_noBKG_WD0708} shows that in the December observations, with no polarised background, the polarisation spectra of WD\,0708$-$670 obtained with background estimated from different strips of the CCD are all fairly consistent among themselves. The polarisation spectra of December are also consistent with the polarisation spectrum obtained in October and calculated using the background in the same strip as the star. Our conclusion is that the background illuminated by the moon does affect the measurement of the polarisation of the star, but that this problem may be minimised by estimating the background in a region very close to the star itself (that is, in the same Wollaston strip). Even though this strategy leads to a lower {\it S/N} than would be achievable if we could estimate the background from a larger area, it minimises the impact of systematic errors. 

It remains to discuss whether the polarisation features may be caused by a internal reflections hitting the spectrum of the star, but sufficiently weak that it appears only during the observations of faint stars and with high background level. We should first note that if these features are due to an internal reflection, then we would expect that its effect varies as the retarder waveplate rotates, and this variable effect would manifest itself as a bump in the null profiles, which we do not see. Furthermore, the same internal scattering should appear in other observations of WDs, while none of the other spectra of our survey (except for WD\,0004$+$122) show similar features. Our conclusion is that the polarisation features seen in the spectrum of WD\,0708$-$670 are real.

The situation for WD\,0004$+$122 is qualitatively similar to the October observations of WD\,0708$-$670;  WD\,0004$+$122 was observed on 2019-10-07 at 90\degr\ from the Moon and with fraction of lunar illumination =0.95. Although we do not have another observation obtained during dark time that confirms our detection, we believe that it is reasonable to assume that at least the observed feature around 6000-7000\,\AA\ is real.

\subsection{ WD\,2049$-$253}
With its nearly constant level of $\sim 0.3-0.5$\,\%, the observed circular polarisation of WD\,2049$-$253 is at the very limit between what could be real and what could be due to instrumental polarisation. The star was observed in dark time, hence there is no significant contamination from background. We also note that if the observed polarisation was due to a cross-talk from linear polarisation, it would probably appear to be wavelength dependent. The observed polarisation could still be due to cross-talk from intensity to circular polarisation. As discussed above, there is no previous evidence for cross-talk from intensity to circular polarisation as large as 0.3\,\%. We therefore conclude that the signal is probably real, but further observations should be obtained (ideally with grism 300V) to confirm our conclusions.

\subsection{WD\,1315$-$781}
We consider our detection of line polarisation in the DA star WD~1315$-$781 quite robust. First, a similar signal has been detected in two independent observations. In both observations, the null profiles oscillate around zero within uncertainties. Moreover, there are no obvious instrumental issues that could mimic Zeeman splitting and polarisation of relatively broad H Balmer lines. Both spectra show a small fraction of polarisation in the continuum, positive ($\sim +0.25$\,\%) on January 8 2020, and negative ($\sim -0.1$\,\%) on January 13. The observations of January 8, 2020 were obtained with very bad seeing, and it was not possible to measure the background from the same slit as the source; the background was not strongly polarised, but very high compared to the source, and we note that the continuum polarisation strongly depend on the area from which the background is calculated. Therefore the reliability of our detection in the continuum is highly questionable. The observations of January 13, 2020 where obtained with a strong polarised background (moon had a fraction of illumination of 92\,\% and was 98\degr\ from the star) and the cross-talk from linear to circular polarisation changes very rapidly with the distance from the star; while more reliable than the observations obtained on January 8, we cannot discard the possibility that also the polarisation measured on January 13 is spurious. Our conclusion is that the nature of the continuum polarisation of WD\,1315$-$781 cannot be assessed from our observations, and was not be considered in our scientific analysis.

\subsection{WD\,1703$-$267}
We consider our detection of line polarisation in the DA star WD\,1703$-$267 very robust. As for the previous case of WD\,1315$-$781, the signal has been detected in two independent observations, and the null profiles oscillate around zero within uncertainties; furthermore, the shape of the Stokes profiles of H Balmer lines corresponds to what we expect from a relatively strong magnetic field. The second spectrum of WD\,1703$-$267 shows a small signal of continuum polarisation. The star was observed in dark time, therefore the background level is low, and not polarised. We do not see an obvious reason why this signal of polarisation should be spurious, although it could be possibly explained by cross-talk from intensity to circular polarisation. Further accurate measurements would be useful to confirm the nature of this signal in the continuum, which we deem as probably intrinsic to the source.

\subsection{WD\,0810$-$353}
The situation for DA star WD\,0810$-$353 is potentially suspicious because the observed polarisation is not supported by a clear physical interpretation; however, the star's unusual polarisation features have been consistently detected in three spectra obtained at two different epochs and with two different grisms. In all observations the null profiles oscillate around zero within photon-noise error bars. Although the observations were obtained with a highly polarised background, the star is bright enough ($V \sim 14.5$) that the stellar spectrum has still a high contrast against the background. The polarisation is concentrated around certain broad bands, but does not appear to have a strong trend in the continuum; in particular it is not present in the blue end of the spectra where the relative contribution of the polarised background is highest. These considerations lead us to the conclusion that all the observed polarisation features of WD\,0810$-$353 are intrinsic to the star.

\section{Discussion}\label{Sect_Discussion}

%%%%%%%%%%%%%%%%%%%%%%%%%%%%%%
%%%%%%%%%%%%%%%%%%%%%%%%%%%%%%%%%%%%%%%%%%%%%%%%%%%%%%%%%%%%%%%%%%%%%%%%%%%%%%%%%%%%%%%%%%%%%%%%%%%%%%%%%%%
\begin{table*}[t]
\caption{MWDs in the \tpv\ in which continuum polarisation is detected, excluding DQ MWDs}
\label{Tab_CP_MWDs}
    \centering
    \begin{tabular}{l l c c c c c c c c c c}
    \hline\hline
\multicolumn{2}{c}{STAR}& $V/G$ & Spec.  & $\pi$ & \te   & $M$      & Age   & \bs  &$V/I$ & References    \\
          &             &      & class & (mas) & (K) & ($M_\circ$)& (Gyr) & (MG) &(\%)         &        \\
    \hline
WD\,0004$+$122 & LP 464--57    & 16.3 & DCH   & 57.3 &  4885& 0.625 & 6.45 & $\sim 200$ & $-0.5$ to +2.0 & 1      \\
WD\,0553+053   & G99--47       & 14.1 & DAH  & 125.0 &  5790& 0.73  & 4.27 & 16         & +0.3 to +0.5   & 2,3,4  \\
WD\,0708$-$670 & SCR J0708-6706& 16.2 & DCH   & 59.0 &  5020& 0.56  & 5.47 & $\sim 200$ & $-0.5$ to +1   & 1      \\
WD\,0810$-$353 &UPM J0812$-$3529& 14.5& DAH   & 89.5 &  6220& 0.69  & 3.1  & $\sim 30$  & $-1$ to +1     & 1      \\
WD\,0912$+$536 & G195--19      & 13.9 & DCH   & 97.3 &  7235& 0.8   & 2.48 & $\sim 100$ & $-2$  to +0.8  & 5,6,7 \\
WD\,1703$-$267 & See text      & 15.0 & DAH   & 76.6 &  6168& 0.82  & 4.2  &  8         &  0.25          & 1  \\
WD\,1748$+$708 & G240--72      & 14.2 & DXH   &161.0 &  5570& 0.8   & 5.86 & $\sim 300$ &$-1.5$ to +1    & 8, 9, 10 \\
WD\,1829$+$547 & G227--35      & 15.5 & DBAH  & 58.7 &  6280& 0.8   & 4.65 & 180        &$-5$ to $-2$    & 11,4,9\\
WD\,1900$+$705 & \grw          & 13.2 & DAH   & 77.7 & 12000& 0.93  & 0.91 & 250        & $-2$  to +6    & 12 to 19 \\
WD\,2049$-$253 & See text      & 16.0 & DC(H?)& 55.7 &  4915& 0.49  & 4.4  &  20        & $0.3$ to 0.5   & 1 \\
    \hline
    \end{tabular}
    \tablefoot{Key to references: 
    (1) this work; 
    (2) \citet{AngeLand72};
    (3) \citet{Liebetal75};
    (4) \citet{PutnJord95}; 
    (5) \citet{AngeLand71}; 
    (6) \citet{Angeetal72b}; 
    (7) \citet{Angeetal81}; 
    (8) \citet{Angeetal74}; 
    (9) \citet{West89}; 
    (10) \citet{BerdPiir99}; 
    (11) \citet{Angeetal75}; 
    (12) \citet{Kempetal70}; 
    (13) \citet{AngeLand70}; 
    (14) \citet{Angeetal72a}; 
    (15) \citet{LandAnge75}; 
    (16) \citet{Angeetal85}; 
    (17) \citet{Jord92}; 
    (18) \citet{Putn95}; 
    (19) \citet{BagnLand19b}. 
    }
\end{table*}

%%%%%%%%%%%%%%%%%%%%%%%%%%%%%%
This work is part of a survey of magnetic fields in the WDs within 20\,pc from the Sun that we have been carrying out with the FORS instrument of the ESO VLT, the ISIS instrument of the WHT, and the ESPaDOnS instrument of the CHFT. Because our efforts come after numerous polarimetric surveys of MWDs (which started 50 years ago), and because magnetic fields of the order of MG or more can be readily detected in classification spectra, many of the new discoveries that we made in the course of our survey were of WDs with comparatively weak magnetic fields. These weak fields could be detected only thanks to the use of large telescopes and modern instrument capable of detecting a subtle Zeeman effect in the circular polarised profiles of spectral lines. The discovery of a significant number of new WDs with strong magnetic fields (from $\sim 5$ to 200\,MG strength) in the \tpv\ is somewhat unexpected, and will have an impact on the statistics of the distribution of field and field strength in WDs.

Four of the six new MWDs presented in this paper were discovered only because we observed them polarimetrically. None of the three DC stars WD\,0004$+$122, WD\,0708$-$670 and WD\,2049$-$253 shows any clear features in Stokes $I$ that identify them as magnetic stars, but all show continuum circular polarisation over at least part of the visible and near IR spectral region. In addition, WD\,0810$-$353, newly identified as a WD, is clearly found to be magnetic by the presence of features in the circular polarisation spectrum, since the features in Stokes~$I$ are probably too subtle to be firmly ascribed to the presence of a magnetic field. Only the two intermediate field MWDs, WD\,1315$-$781 and WD\,1703$-$267, have easily recognisable Zeeman splitting in Stokes~$I$, around \ha\ and H$\beta$, which are confirmed in both cases by Zeeman polarisation features coinciding with the $I$ Zeeman components of the Balmer lines. The analysis of these new MWDs highlights the importance of WDs as benchmarks for atomic physics, in particular for the study of line formation and radiative transfer in the presence of a magnetic field well beyond the familiar linear Zeeman regime.

MWDs with fields strength of several tens of MG or more have often been identified as such by polarimetric measurements, because their intensity spectra may lack recognisable lines. The number of MWDs within the \tpv\ that are known to show non-zero continuum circular polarisation (excluding the DQ MWDs, which seem to have very distinctive polarisation properties) has increased from five to ten members, listed in Table\,\ref{Tab_CP_MWDs}. In this table, the characteristic field value \bs\ represents in some cases only a very rough estimate; the range of $V/I$ tabulated describes approximately the range of continuum circular polarisation exclusive of strong excursions due to spectral lines. 

It is notable that the stars within the \tpv\ that show detected continuum circular polarisation are systematically different from the sample of similar stars lying at greater distances. All but two of the MWDs in Table\,\ref{Tab_CP_MWDs} have $\te \la 6500$\,K, while the more distant MWDs of this type generally have $\te \ga 6500$\,K, with values ranging up to 30\,000\,K and more. This situation is well known, and is an effect of an observational bias outside of the \tpv, where cooler (hence fainter) stars have probably been observed less frequently than hotter ones. The local sample provides a nearly unique view of the oldest MWDs showing continuum polarisation, which are often the MWDs with the strongest fields. The five new stars double the length of this particular list, and three of the five new stars are the coolest stars in the \tpv\ continuum-polarised sample that are not DQpecP MWDs. 

The main reason for including Table\,\ref{Tab_CP_MWDs} in this paper is to highlight an important kind of challenge represented by cool, old, very strong field MWDs. It is possible to derive plausible models of the surface magnetic field structures of many MWDs with weak (sub-MG) and intermediate (few MG and tens of MG) fields, exploiting the spectral line structures seen in both $I$ and $V/I$ spectra \citep[e.g.][]{AchiWick89,PutnJord95,Landetal17}. The challenge of deriving magnetic surface models and maps is much more severe for fields from roughly 100\,MG on up, and only a very small number of such models have so far been produced \citep[e.g.][]{WickFerr88,Jord92,Euchetal06}. This challenge is even greater for the coolest stars with large fields, and the models of the highest field MWDs are all for hotter MWDs than those discussed here.

 Table\,\ref{Tab_CP_MWDs} points up the fact that an important population of high-field MWDs exists which are old stars of low \te. These stars make up perhaps one quarter of the MWDs in the \tpv. Most are stars 
 for which the information in the $I$ spectra alone is not sufficient to permit detailed modelling. We have shown that valuable constraints on the fields present in some of these stars may be obtainable from weak spectral line-like features in the $V/I$ spectra. In addition, the wavelength dependence of the overall broad-band polarisation remains to be explained and exploited. The magnitude of this latter challenge may be indicated by the successful model of the circular polarisation spectrum of \grw\ (a star having a rich $I$ spectrum with which to constrain models) developed by \citet{Jord03}, in which success was achieved by representing the magnetic field with a 25-parameter expansion! The stars of Table\,\ref{Tab_CP_MWDs} contain important information about the long-time evolution of the strongest magnetic fields, but we are still only beginning even to be able to characterise the fields of such stars.

\section{Conclusions}\label{Sect_Conclusions}

We have reported the discovery of six previously unknown MWDs that are within the \tpv. Three of these newly discovered MWDs (WD\,0810-353,  WD\,1703$-$267, and WD\,2049-253) are stars that were only very recently identified as WDs and as members of the \tpv, thanks to the USNO southern parallax programme and to large-scale data mining in the treasure trove of Gaia DR2; the three other are well-known WDs that have previously been studied.
 
 Two of the new MWDs are clearly DA WDs. WD\,1315$-$78 was thought to be of spectral type DC because the previously available spectroscopy did not have high enough {\it S/N} to reliably demonstrate the presence of weak, Zeeman split, \ha\ and H$\beta$ triplets.  WD\,1703$-$267 is newly identified as a WD and member of the \tpv, without previous spectroscopy. The features of the intensity spectrum of a third new MWD (WD\,0810$-$353) are so weak to be barely distinguishable from noise and imperfect flat-fielding correction. Line-like Stokes~$V$ profiles are very clearly detected, but only tentatively ascribed to H$\beta$. The star is probably a DA MWD with Stokes~$I$ spectral lines washed out by a strong magnetic field. Further observations around H$\alpha$ may help to clarify the nature of this star.

 Two of the remaining three new magnetic stars (WD\,0004$+$122 and WD\,0708$-$670) were known DC WDs that had never been examined with any kind of circular polarisation measurement, because polarimetric surveys of such stars are very incomplete at the $V \sim 16$ magnitude level, especially in the southern sky. The third new DCH star, WD\,2049-253, was only recently identified as a member of the \tpv, and has no previous spectroscopy.

The three new DA MWDs exhibit spectral lines remarkably offset and split by the presence of a strong magnetic field. For two of them we have been able to quantitatively interpret the spectra and estimate the field strength. Although we have a preliminary, qualitative interpretation of the spectrum of a third MWD, WD\,0810$-$353, this star requires further observational and theoretical work.

The three new DC MWDs provide examples of magnetism in the old and cool end of the evolution sequence of stars that are not members of the DA or DQ class. The ease with which they were found, and the relative shortage of high-field MWDs south of the celestial equator, suggests that there may be more such stars to be identified among the DC stars not so far surveyed with spectropolarimetry. The state of modelling of such cool, strongly magnetic WDs, for which we have no useful information from the Stokes~$I$ spectrum, is still particularly underdeveloped, although it is now clear that such stars represent a significant fraction of the full sample of MWDs. 

To estimate of the field strength of these new DC MWDs we have worked out an empirical relationship between the level of polarisation of the continuum in cool MWDs and field strength, using observations of stars that show both polarisation in the continuum and in spectral lines. We have estimated that a 15\,MG longitudinal field produces $\sim 1$\,\% of polarisation in the continuum. In the course of this investigation we have also tried to clarify the various ambiguities around the definition of the sign of circular polarisation and its relationship with the sign of the longitudinal component of the magnetic field, and we have shown that a positive field produces a right handed circular polarisation (as seen from the observer).
 
The incidence of MWDs has been previously studied, for example by \citet{Kawketal07}. From the population of WDs within 13\,pc  they estimated that $21\,\pm 8\,$\% of WDs are magnetic, and from the WD population in the local 20\,pc volume, they  estimated a frequency of $13\pm 4$\,\%. From more recent observations and analysis of the local \tpv\, \citet{LandBagn19b} found that the frequency of DA-type MWDs in the \tpv\ is $20 \pm 5$\,\%. Our new discoveries increase the number of known MWDs in the \tpv\ by a further $\sim 20$\,\%, and the fraction of MWDs among the approximately 150 WDs of the same volume by about 4\,\%. In the same volume, the number of known large-field (50\,MG or more) MWDs is increased by about 35\,\%. In summary, we have presented a group of new entrants that is large enough to significantly alter all the statistics of frequency of MWDs. These statistics will be discussed in greater detail in a forthcoming paper when we our survey is completed. 

One of the very important lessons to take away from our results is the ongoing importance of spectropolarimetry in finding and studying new MWDs, especially old, cool and faint WDs. Only two of the six new MWDs could easily have been identified through typical spectroscopic survey observations, even though all of them have fields of at least several MG. At the same time, the difficulties that we have encountered in the analysis of the data of faint stars observed in presence of a bright linear polarised background underscore the need for a full characterisation of the instruments used for this kind of surveys.

\begin{acknowledgements}
We thank the anonymous referee for a very thorough reading of the manuscript and for a number of valuable comments and suggestions. This work is based on observations made with ESO Telescopes at the La Silla Paranal Observatory under programme IDs 0103.D-0029 and 0104.D-0298. JDL acknowledges financial support from the Natural Sciences and Engineering Research Council of Canada, (NSERC), funding reference number 6377-2016.
\end{acknowledgements}

\bibliography{jdl-sb}

\begin{thebibliography}{99}
\expandafter\ifx\csname natexlab\endcsname\relax\def\natexlab#1{#1}\fi

\bibitem[{{Achilleos} \& {Wickramasinghe}(1989)}]{AchiWick89}
{Achilleos}, N. \& {Wickramasinghe}, D.~T. 1989, \apj, 346, 444

\bibitem[{{Angel} {et~al.}(1981){Angel}, {Borra}, \& {Landstreet}}]{Angeetal81}
{Angel}, J.~R.~P., {Borra}, E.~F., \& {Landstreet}, J.~D. 1981, \apjs, 45, 457

\bibitem[{{Angel} {et~al.}(1974{\natexlab{a}}){Angel}, {Carswell},
  {Strittmatter}, {Beaver}, \& {Harms}}]{Angeetal74b}
{Angel}, J.~R.~P., {Carswell}, R.~F., {Strittmatter}, P.~A., {Beaver}, E.~A.,
  \& {Harms}, R. 1974{\natexlab{a}}, \apjl, 194, L47

\bibitem[{{Angel} {et~al.}(1975){Angel}, {Hintzen}, \&
  {Landstreet}}]{Angeetal75}
{Angel}, J.~R.~P., {Hintzen}, P., \& {Landstreet}, J.~D. 1975, \apjl, 196, L27

\bibitem[{{Angel} {et~al.}(1974{\natexlab{b}}){Angel}, {Hintzen},
  {Strittmatter}, \& {Martin}}]{Angeetal74}
{Angel}, J.~R.~P., {Hintzen}, P., {Strittmatter}, P.~A., \& {Martin}, P.~G.
  1974{\natexlab{b}}, \apjl, 190, L71

\bibitem[{{Angel} {et~al.}(1972{\natexlab{a}}){Angel}, {Illing}, \&
  {Landstreet}}]{Angeetal72b}
{Angel}, J.~R.~P., {Illing}, R.~M.~E., \& {Landstreet}, J.~D.
  1972{\natexlab{a}}, \apjl, 175, L85

\bibitem[{{Angel} \& {Landstreet}(1970)}]{AngeLand70}
{Angel}, J.~R.~P. \& {Landstreet}, J.~D. 1970, \apjl, 162, L61

\bibitem[{{Angel} \& {Landstreet}(1971)}]{AngeLand71}
{Angel}, J.~R.~P. \& {Landstreet}, J.~D. 1971, \apjl, 165, L71

\bibitem[{{Angel} \& {Landstreet}(1972)}]{AngeLand72}
{Angel}, J.~R.~P. \& {Landstreet}, J.~D. 1972, \apjl, 178, L21

\bibitem[{{Angel} \& {Landstreet}(1974)}]{AngeLand74}
{Angel}, J.~R.~P. \& {Landstreet}, J.~D. 1974, \apj, 191, 457

\bibitem[{{Angel} {et~al.}(1972{\natexlab{b}}){Angel}, {Landstreet}, \&
  {Oke}}]{Angeetal72a}
{Angel}, J.~R.~P., {Landstreet}, J.~D., \& {Oke}, J.~B. 1972{\natexlab{b}},
  \apjl, 171, L11

\bibitem[{{Angel} {et~al.}(1985){Angel}, {Liebert}, \& {Stockman}}]{Angeetal85}
{Angel}, J.~R.~P., {Liebert}, J., \& {Stockman}, H.~S. 1985, \apj, 292, 260

\bibitem[{{Aznar Cuadrado} {et~al.}(2004){Aznar Cuadrado}, {Jordan},
  {Napiwotzki}, {Schmid}, {Solanki}, \& {Mathys}}]{Aznaetal04}
{Aznar Cuadrado}, R., {Jordan}, S., {Napiwotzki}, R., {et~al.} 2004, \aap, 423,
  1081

\bibitem[{{Bagnulo} {et~al.}(2017){Bagnulo}, {Cox}, {Cikota}, {Siebenmorgen},
  {Voshchinnikov}, {Patat}, {Smith}, {Smoker}, {Taubenberger}, {Kaper}, {Cami},
  \& {LIPS Collaboration}}]{Bagetal17}
{Bagnulo}, S., {Cox}, N. L.~J., {Cikota}, A., {et~al.} 2017, \aap, 608, A146

\bibitem[{{Bagnulo} {et~al.}(2013){Bagnulo}, {Fossati}, {Kochukhov}, \&
  {Landstreet}}]{Bagetal13}
{Bagnulo}, S., {Fossati}, L., {Kochukhov}, O., \& {Landstreet}, J.~D. 2013,
  \aap, 559, A103

\bibitem[{{Bagnulo} {et~al.}(2009){Bagnulo}, {Landolfi}, {Landstreet}, {Land i
  Degl'Innocenti}, {Fossati}, \& {Sterzik}}]{Bagetal09}
{Bagnulo}, S., {Landolfi}, M., {Landstreet}, J.~D., {et~al.} 2009, \pasp, 121,
  993

\bibitem[{{Bagnulo} \& {Landstreet}(2018)}]{BagnLand18}
{Bagnulo}, S. \& {Landstreet}, J.~D. 2018, \aap, 618, A113

\bibitem[{{Bagnulo} \& {Landstreet}(2019{\natexlab{a}})}]{BagnLand19b}
{Bagnulo}, S. \& {Landstreet}, J.~D. 2019{\natexlab{a}}, \aap, 630, A65

\bibitem[{{Bagnulo} \& {Landstreet}(2019{\natexlab{b}})}]{BagnLand19a}
{Bagnulo}, S. \& {Landstreet}, J.~D. 2019{\natexlab{b}}, \mnras, 486, 4655

\bibitem[{{Bagnulo} {et~al.}(2012){Bagnulo}, {Landstreet}, {Fossati}, \&
  {Kochukhov}}]{Bagetal12}
{Bagnulo}, S., {Landstreet}, J.~D., {Fossati}, L., \& {Kochukhov}, O. 2012,
  \aap, 538, A129

\bibitem[{{Berdyugin} \& {Piirola}(1999)}]{BerdPiir99}
{Berdyugin}, A.~V. \& {Piirola}, V. 1999, \aap, 352, 619

\bibitem[{{Berdyugina} {et~al.}(2007){Berdyugina}, {Berdyugin}, \&
  {Piirola}}]{Berdetal07}
{Berdyugina}, S.~V., {Berdyugin}, A.~V., \& {Piirola}, V. 2007, \prl, 99,
  091101

\bibitem[{{Blouin} {et~al.}(2019){Blouin}, {Dufour}, {Thibeault}, \&
  {Allard}}]{Blouetal19}
{Blouin}, S., {Dufour}, P., {Thibeault}, C., \& {Allard}, N.~F. 2019, \apj,
  878, 63

\bibitem[{{Borra} \& {Landstreet}(1980)}]{BorLan80}
{Borra}, E.~F. \& {Landstreet}, J.~D. 1980, \apjs, 42, 421

\bibitem[{{Cantiello} {et~al.}(2016){Cantiello}, {Fuller}, \&
  {Bildsten}}]{Cantetal16}
{Cantiello}, M., {Fuller}, J., \& {Bildsten}, L. 2016, \apj, 824, 14

\bibitem[{{Chiuderi} \& {Silvi}(1976)}]{ChiSil76}
{Chiuderi}, C. \& {Silvi}, M. 1976, \memsai, 47, 65

\bibitem[{{Cohen} {et~al.}(1993){Cohen}, {Putney}, \& {Goodrich}}]{Coheetal93}
{Cohen}, M.~H., {Putney}, A., \& {Goodrich}, R.~W. 1993, \apjl, 405, L67

\bibitem[{{Cummings} {et~al.}(2018){Cummings}, {Kalirai}, {Tremblay},
  {Ramirez-Ruiz}, \& {Choi}}]{Cumetal18}
{Cummings}, J.~D., {Kalirai}, J.~S., {Tremblay}, P.~E., {Ramirez-Ruiz}, E., \&
  {Choi}, J. 2018, \apj, 866, 21

\bibitem[{{Donati} {et~al.}(1997){Donati}, {Semel}, {Carter}, {Rees}, \&
  {Collier Cameron}}]{Donaetal97}
{Donati}, J.~F., {Semel}, M., {Carter}, B.~D., {Rees}, D.~E., \& {Collier
  Cameron}, A. 1997, \mnras, 291, 658

\bibitem[{{Euchner} {et~al.}(2006){Euchner}, {Jordan}, {Beuermann}, {Reinsch},
  \& {G{\"a}nsicke}}]{Euchetal06}
{Euchner}, F., {Jordan}, S., {Beuermann}, K., {Reinsch}, K., \& {G{\"a}nsicke},
  B.~T. 2006, \aap, 451, 671

\bibitem[{{Ferrario} {et~al.}(2015){Ferrario}, {de Martino}, \&
  {G{\"a}nsicke}}]{Ferretal15}
{Ferrario}, L., {de Martino}, D., \& {G{\"a}nsicke}, B.~T. 2015, \ssr, 191, 111

\bibitem[{{Finch} {et~al.}(2007){Finch}, {Henry}, {Subasavage}, {Jao}, \&
  {Hambly}}]{Fincetal07}
{Finch}, C.~T., {Henry}, T.~J., {Subasavage}, J.~P., {Jao}, W.-C., \& {Hambly},
  N.~C. 2007, \aj, 133, 2898

\bibitem[{{Finch} {et~al.}(2018){Finch}, {Zacharias}, \& {Jao}}]{Fincetal18}
{Finch}, C.~T., {Zacharias}, N., \& {Jao}, W.-C. 2018, \aj, 155, 176

\bibitem[{{Fontaine} {et~al.}(1973){Fontaine}, {Thomas}, \& {van
  Horn}}]{Fontetal73}
{Fontaine}, G., {Thomas}, J.~H., \& {van Horn}, H.~M. 1973, \apj, 184, 911

\bibitem[{{Forster} {et~al.}(1984){Forster}, {Strupat}, {Rosner}, {Wunner},
  {Ruder}, \& {Herold}}]{Forsetal84}
{Forster}, H., {Strupat}, W., {Rosner}, W., {et~al.} 1984, Journal of Physics B
  Atomic Molecular Physics, 17, 1301

\bibitem[{{Gaia Collaboration} {et~al.}(2018){Gaia Collaboration}, {Brown},
  {Vallenari}, {Prusti}, {de Bruijne}, {Babusiaux}, {Bailer-Jones}, {Biermann},
  {Evans}, \& {Eyer}}]{Gaia18}
{Gaia Collaboration}, {Brown}, A.~G.~A., {Vallenari}, A., {et~al.} 2018, \aap,
  616, A1

\bibitem[{{Garc{\'\i}a-Berro} {et~al.}(2012){Garc{\'\i}a-Berro},
  {Lor{\'e}n-Aguilar}, {Aznar-Sigu{\'a}n}, {Torres}, {Camacho}, {Althaus},
  {C{\'o}rsico}, {K{\"u}lebi}, \& {Isern}}]{Garcetal12}
{Garc{\'\i}a-Berro}, E., {Lor{\'e}n-Aguilar}, P., {Aznar-Sigu{\'a}n}, G.,
  {et~al.} 2012, \apj, 749, 25

\bibitem[{{Garstang} \& {Kemic}(1974)}]{GarsKemi74}
{Garstang}, R.~H. \& {Kemic}, S.~B. 1974, \apss, 31, 103

\bibitem[{{Gentile Fusillo} {et~al.}(2019){Gentile Fusillo}, {Tremblay},
  {G{\"a}nsicke}, {Manser}, {Cunningham}, {Cukanovaite}, {Hollands}, {Marsh},
  {Raddi}, {Jordan}, {Toonen}, {Geier}, {Barstow}, \& {Cummings}}]{Gentetal19}
{Gentile Fusillo}, N.~P., {Tremblay}, P.-E., {G{\"a}nsicke}, B.~T., {et~al.}
  2019, \mnras, 482, 4570

\bibitem[{{Giammichele} {et~al.}(2012){Giammichele}, {Bergeron}, \&
  {Dufour}}]{Giametal12}
{Giammichele}, N., {Bergeron}, P., \& {Dufour}, P. 2012, \apjs, 199, 29

\bibitem[{{Henry} \& {O'Connell}(1985)}]{HenrOcon85}
{Henry}, R.~J.~W. \& {O'Connell}, R.~F. 1985, \pasp, 97, 333

\bibitem[{{Hensberge} {et~al.}(1977){Hensberge}, {van Rensbergen}, {Goossens},
  \& {Deridder}}]{Hensetal77}
{Hensberge}, H., {van Rensbergen}, W., {Goossens}, M., \& {Deridder}, G. 1977,
  \aap, 61, 235

\bibitem[{{Holberg} {et~al.}(2016){Holberg}, {Oswalt}, {Sion}, \&
  {McCook}}]{Holbetal16}
{Holberg}, J.~B., {Oswalt}, T.~D., {Sion}, E.~M., \& {McCook}, G.~P. 2016,
  \mnras, 462, 2295

\bibitem[{{Hollands} {et~al.}(2018){Hollands}, {Tremblay}, {G{\"a}nsicke},
  {Gentile-Fusillo}, \& {Toonen}}]{Holletal18}
{Hollands}, M.~A., {Tremblay}, P.~E., {G{\"a}nsicke}, B.~T., {Gentile-Fusillo},
  N.~P., \& {Toonen}, S. 2018, \mnras, 480, 3942

\bibitem[{{Isern} {et~al.}(2017){Isern}, {Garc{\'\i}a-Berro}, {K{\"u}lebi}, \&
  {Lor{\'e}n-Aguilar}}]{Iseretal17}
{Isern}, J., {Garc{\'\i}a-Berro}, E., {K{\"u}lebi}, B., \& {Lor{\'e}n-Aguilar},
  P. 2017, \apjl, 836, L28

\bibitem[{{Jordan}(1992)}]{Jord92}
{Jordan}, S. 1992, \aap, 265, 570

\bibitem[{{Jordan}(2003)}]{Jord03}
{Jordan}, S. 2003, in NATO ASIB Proc. 105: White Dwarfs, Vol. 105, 175

\bibitem[{{Jordan} {et~al.}(2007){Jordan}, {Aznar Cuadrado}, {Napiwotzki},
  {Schmid}, \& {Solanki}}]{Joretal07}
{Jordan}, S., {Aznar Cuadrado}, R., {Napiwotzki}, R., {Schmid}, H.~M., \&
  {Solanki}, S.~K. 2007, \aap, 462, 1097

\bibitem[{{Kawka} \& {Vennes}(2006)}]{KawkVenn06}
{Kawka}, A. \& {Vennes}, S. 2006, \apj, 643, 402

\bibitem[{{Kawka} \& {Vennes}(2014)}]{KawkVenn14}
{Kawka}, A. \& {Vennes}, S. 2014, \mnras, 439, L90

\bibitem[{{Kawka} {et~al.}(2007){Kawka}, {Vennes}, {Schmidt}, {Wickramasinghe},
  \& {Koch}}]{Kawketal07}
{Kawka}, A., {Vennes}, S., {Schmidt}, G.~D., {Wickramasinghe}, D.~T., \&
  {Koch}, R. 2007, \apj, 654, 499

\bibitem[{{Kemp}(1970)}]{Kemp70}
{Kemp}, J.~C. 1970, \apj, 162, 169

\bibitem[{{Kemp}(1977)}]{Kemp77}
{Kemp}, J.~C. 1977, \apj, 213, 794

\bibitem[{{Kemp} {et~al.}(1970{\natexlab{a}}){Kemp}, {Swedlund}, \&
  {Evans}}]{Kempetal70a}
{Kemp}, J.~C., {Swedlund}, J.~B., \& {Evans}, B.~D. 1970{\natexlab{a}}, \prl,
  24, 1211

\bibitem[{{Kemp} {et~al.}(1970{\natexlab{b}}){Kemp}, {Swedlund}, {Landstreet},
  \& {Angel}}]{Kempetal70}
{Kemp}, J.~C., {Swedlund}, J.~B., {Landstreet}, J.~D., \& {Angel}, J.~R.~P.
  1970{\natexlab{b}}, \apjl, 161, L77

\bibitem[{{Kepler} {et~al.}(2013){Kepler}, {Pelisoli}, {Jordan}, {Kleinman},
  {Koester}, {K{\"u}lebi}, {Pe{\c{c}}anha}, {Castanheira}, {Nitta}, {Costa},
  {Winget}, {Kanaan}, \& {Fraga}}]{Kepletal13}
{Kepler}, S.~O., {Pelisoli}, I., {Jordan}, S., {et~al.} 2013, \mnras, 429, 2934

\bibitem[{{Koester} {et~al.}(2009){Koester}, {Voss}, {Napiwotzki},
  {Christlieb}, {Homeier}, {Lisker}, {Reimers}, \& {Heber}}]{Koesetal09}
{Koester}, D., {Voss}, B., {Napiwotzki}, R., {et~al.} 2009, \aap, 505, 441

\bibitem[{{Landi Degl'Innocenti} \& {Landolfi}(2004)}]{LanLan04}
{Landi Degl'Innocenti}, E. \& {Landolfi}, M. 2004, {Polarization in Spectral
  Lines}, Vol. 307

\bibitem[{{Landstreet}(2020)}]{Land20}
{Landstreet}, J.~D. 2020, arXiv e-prints, arXiv:2008.01802

\bibitem[{{Landstreet} \& {Angel}(1971)}]{LandAnge71}
{Landstreet}, J.~D. \& {Angel}, J.~R.~P. 1971, \apjl, 165, L67

\bibitem[{{Landstreet} \& {Angel}(1975)}]{LandAnge75}
{Landstreet}, J.~D. \& {Angel}, J.~R.~P. 1975, \apj, 196, 819

\bibitem[{{Landstreet} \& {Bagnulo}(2019{\natexlab{a}})}]{LandBagn19a}
{Landstreet}, J.~D. \& {Bagnulo}, S. 2019{\natexlab{a}}, \aap, 628, A1

\bibitem[{{Landstreet} \& {Bagnulo}(2019{\natexlab{b}})}]{LandBagn19b}
{Landstreet}, J.~D. \& {Bagnulo}, S. 2019{\natexlab{b}}, \aap, 623, A46

\bibitem[{{Landstreet} {et~al.}(2016){Landstreet}, {Bagnulo}, {Martin}, \&
  {Valyavin}}]{Landetal16}
{Landstreet}, J.~D., {Bagnulo}, S., {Martin}, A., \& {Valyavin}, G. 2016, \aap,
  591, A80

\bibitem[{{Landstreet} {et~al.}(2017){Landstreet}, {Bagnulo}, {Valyavin}, \&
  {Valeev}}]{Landetal17}
{Landstreet}, J.~D., {Bagnulo}, S., {Valyavin}, G., \& {Valeev}, A.~F. 2017,
  \aap, 607, A92

\bibitem[{{Landstreet} {et~al.}(2012){Landstreet}, {Bagnulo}, {Valyavin},
  {Fossati}, {Jordan}, {Monin}, \& {Wade}}]{Landetal12}
{Landstreet}, J.~D., {Bagnulo}, S., {Valyavin}, G.~G., {et~al.} 2012, \aap,
  545, A30

\bibitem[{{Landstreet} {et~al.}(2015){Landstreet}, {Bagnulo}, {Valyavin},
  {Gadelshin}, {Martin}, {Galazutdinov}, \& {Semenko}}]{Landetal15}
{Landstreet}, J.~D., {Bagnulo}, S., {Valyavin}, G.~G., {et~al.} 2015, \aap,
  580, A120

\bibitem[{{Leroy}(1999)}]{Lero99}
{Leroy}, J.~L. 1999, \aap, 346, 955

\bibitem[{{Liebert} {et~al.}(1975){Liebert}, {Angel}, \&
  {Landstreet}}]{Liebetal75}
{Liebert}, J., {Angel}, J.~R.~P., \& {Landstreet}, J.~D. 1975, \apjl, 202, L139

\bibitem[{{Liebert} {et~al.}(1977){Liebert}, {Angel}, {Stockman}, {Spinrad}, \&
  {Beaver}}]{Liebetal77}
{Liebert}, J., {Angel}, J.~R.~P., {Stockman}, H.~S., {Spinrad}, H., \&
  {Beaver}, E.~A. 1977, \apj, 214, 457

\bibitem[{{Limoges} {et~al.}(2015){Limoges}, {Bergeron}, \&
  {L{\'e}pine}}]{Limoetal15}
{Limoges}, M.~M., {Bergeron}, P., \& {L{\'e}pine}, S. 2015, \apjs, 219, 19

\bibitem[{{Luyten}(1949)}]{Luyt49}
{Luyten}, W.~J. 1949, \apj, 109, 528

\bibitem[{{Martin} \& {Wickramasinghe}(1979)}]{MartWick79}
{Martin}, B. \& {Wickramasinghe}, D.~T. 1979, \mnras, 189, 883

\bibitem[{{Mathys}(1989)}]{Math89}
{Mathys}, G. 1989, \fcp, 13, 143

\bibitem[{{Mestel}(1952)}]{Mest52}
{Mestel}, L. 1952, \mnras, 112, 583

\bibitem[{{Putney}(1995)}]{Putn95}
{Putney}, A. 1995, \apjl, 451, L67

\bibitem[{{Putney}(1997)}]{Putn97}
{Putney}, A. 1997, \apjs, 112, 527

\bibitem[{{Putney} \& {Jordan}(1995)}]{PutnJord95}
{Putney}, A. \& {Jordan}, S. 1995, \apj, 449, 863

\bibitem[{{Roesner} {et~al.}(1984){Roesner}, {Wunner}, {Herold}, \&
  {Ruder}}]{Roesetal84}
{Roesner}, W., {Wunner}, G., {Herold}, H., \& {Ruder}, H. 1984, Journal of
  Physics B Atomic Molecular Physics, 17, 29

\bibitem[{{Salim} \& {Gould}(2003)}]{SalGou03}
{Salim}, S. \& {Gould}, A. 2003, \apj, 582, 1011

\bibitem[{{Schimeczek} \& {Wunner}(2014)}]{SchiWunn14}
{Schimeczek}, C. \& {Wunner}, G. 2014, \apjs, 212, 26

\bibitem[{{Schmidt} \& {Norsworthy}(1991)}]{SchmNors91}
{Schmidt}, G.~D. \& {Norsworthy}, J.~E. 1991, \apj, 366, 270

\bibitem[{{Schmidt} \& {Smith}(1995)}]{SchmSmit95}
{Schmidt}, G.~D. \& {Smith}, P.~S. 1995, \apj, 448, 305

\bibitem[{{Shipman}(1971)}]{Ship71}
{Shipman}, H.~L. 1971, \apj, 167, 165

\bibitem[{{Siebenmorgen} {et~al.}(2014){Siebenmorgen}, {Voshchinnikov}, \&
  {Bagnulo}}]{Sieetal14}
{Siebenmorgen}, R., {Voshchinnikov}, N.~V., \& {Bagnulo}, S. 2014, \aap, 561,
  A82

\bibitem[{{Subasavage} {et~al.}(2008){Subasavage}, {Henry}, {Bergeron},
  {Dufour}, \& {Hambly}}]{Subaetal08}
{Subasavage}, J.~P., {Henry}, T.~J., {Bergeron}, P., {Dufour}, P., \& {Hambly},
  N.~C. 2008, \aj, 136, 899

\bibitem[{{Subasavage} {et~al.}(2007){Subasavage}, {Henry}, {Bergeron},
  {Dufour}, {Hambly}, \& {Beaulieu}}]{Subaetal07}
{Subasavage}, J.~P., {Henry}, T.~J., {Bergeron}, P., {et~al.} 2007, \aj, 134,
  252

\bibitem[{{Subasavage} {et~al.}(2017){Subasavage}, {Jao}, {Henry}, {Harris},
  {Dahn}, {Bergeron}, {Dufour}, {Dunlap}, {Barlow}, {Ianna}, {L{\'e}pine}, \&
  {Margheim}}]{Subaetal17}
{Subasavage}, J.~P., {Jao}, W.-C., {Henry}, T.~J., {et~al.} 2017, \aj, 154, 32

\bibitem[{{Swedlund} {et~al.}(1974){Swedlund}, {Wolstencroft}, {Michalsky}, \&
  {Kemp}}]{Swedetal74}
{Swedlund}, J.~B., {Wolstencroft}, R.~D., {Michalsky}, Joseph~J., J., \&
  {Kemp}, J.~C. 1974, \apjl, 187, L121

\bibitem[{{Tout} {et~al.}(2008){Tout}, {Wickramasinghe}, {Liebert}, {Ferrario},
  \& {Pringle}}]{Toutetal08}
{Tout}, C.~A., {Wickramasinghe}, D.~T., {Liebert}, J., {Ferrario}, L., \&
  {Pringle}, J.~E. 2008, \mnras, 387, 897

\bibitem[{{Tremblay} {et~al.}(2011){Tremblay}, {Bergeron}, \&
  {Gianninas}}]{Tremetal11}
{Tremblay}, P.~E., {Bergeron}, P., \& {Gianninas}, A. 2011, \apj, 730, 128

\bibitem[{{Vornanen} {et~al.}(2010){Vornanen}, {Berdyugina}, {Berdyugin}, \&
  {Piirola}}]{Vornetal10}
{Vornanen}, T., {Berdyugina}, S.~V., {Berdyugin}, A.~V., \& {Piirola}, V. 2010,
  \apjl, 720, L52

\bibitem[{{Wade} {et~al.}(2001){Wade}, {Bagnulo}, {Kochukhov}, {Land street},
  {Piskunov}, \& {Stift}}]{Wadetal01}
{Wade}, G.~A., {Bagnulo}, S., {Kochukhov}, O., {et~al.} 2001, \aap, 374, 265

\bibitem[{{Wade} {et~al.}(2000){Wade}, {Donati}, {Landstreet}, \&
  {Shorlin}}]{Wadetal00a}
{Wade}, G.~A., {Donati}, J.~F., {Landstreet}, J.~D., \& {Shorlin}, S.~L.~S.
  2000, \mnras, 313, 823

\bibitem[{{West}(1989)}]{West89}
{West}, S.~C. 1989, \apj, 345, 511

\bibitem[{{Wickramasinghe} \& {Bessell}(1976)}]{WickBess76}
{Wickramasinghe}, D.~T. \& {Bessell}, M.~S. 1976, \apjl, 203, L39

\bibitem[{{Wickramasinghe} \& {Cropper}(1988)}]{WickCrop88}
{Wickramasinghe}, D.~T. \& {Cropper}, M. 1988, \mnras, 235, 1451

\bibitem[{{Wickramasinghe} \& {Ferrario}(1988)}]{WickFerr88}
{Wickramasinghe}, D.~T. \& {Ferrario}, L. 1988, \apj, 327, 222

\bibitem[{{Wunner} {et~al.}(1985){Wunner}, {Roesner}, {Herold}, \&
  {Ruder}}]{Wunnetal85}
{Wunner}, G., {Roesner}, W., {Herold}, H., \& {Ruder}, H. 1985, \aap, 149, 102

\end{thebibliography}

\end{document}